\documentclass[useAMS,usenatbib]{mn2e}
\usepackage{psfig,epsf}
\begin{document}

\title[The cored distribution of dark matter in spiral galaxies.]
{The cored distribution of dark matter in spiral galaxies\\}
\author [G. Gentile, P. Salucci, U. Klein, D. Vergani and P. Kalberla]
{G. Gentile $^{1}$\thanks{E-mail: ggentile@astro.uni-bonn.de}, 
P. Salucci$^{2}$, U. Klein$^{1}$, D. Vergani $^{1,3}$ and P. Kalberla$^{1}$\\
$^{1}$ Radioastronomisches Institut der Universit\"at Bonn,
Auf dem H\"ugel 71, 53121 Bonn, Germany \\
$^{2}$ SISSA, via Beirut 4, 34014 Trieste, Italy \\
$^{3}$ Observatoire de Paris, GEPI, CNRS UMR 8111 and 
Universit\'e Paris 7, 5 place Jules Janssen, 
92195 Meudon Cedex, France}

\date{Accepted  Received}

\maketitle

\begin{abstract}
We present the {H{\sc i}} data for 5 spiral galaxies 
that, along with their H$\alpha$ rotation curves, are used 
to derive the distribution of dark matter within these objects. A new method for extracting
rotation curves from {H{\sc i}} data cubes is presented; this takes into 
account the existence of a warp and 
minimises projection effects.
The rotation curves obtained are tested by taking them as input to 
construct model data cubes that are compared to the observed ones: the 
agreement is excellent.
On the contrary, the model data cubes built using
rotation curves obtained with standard methods, such as the first-moment analysis,
fail the test.
The {H{\sc i}} rotation curves agree well with the H$\alpha$ data, where they coexist.
Moreover, the combined H$\alpha$ + {H{\sc i}} rotation curves  
are smooth, symmetric and 
extended to large radii.

The rotation curves are decomposed into stellar, gaseous
and dark matter contributions and the inferred density distribution 
is compared to various mass distributions:
dark haloes with a central density core, $\Lambda$ Cold Dark Matter ($\Lambda$CDM) haloes
(NFW, Moore profiles), {H{\sc i}} scaling and MOND.
The observations point to
haloes with constant density cores of size $r_{core} \sim r_{opt}$ and
central densities scaling approximately as $\rho_0 \propto r_{core}^{-2/3}$.
$\Lambda$CDM models (which predict a central cusp in the density
profile) are in clear conflict with the data. 
{H{\sc i}} scaling
and MOND cannot account for the observed kinematics:
we find some counter-examples.

\end{abstract}


\section{Introduction}

One of the tantalising and yet unanswered questions of contemporary 
astrophysics is that concerning the nature of dark matter (DM). While its existence
has been inferred for several decades, only in the recent
past its amount has been carefully assessed on various astrophysical scales.
On galaxy scales, the classical and still most powerful tool to unveil
DM is that of rotation curves. Compared to the pioneering work of \citet*{Ru:80}
and \citet{Bo:81}, modern studies take advantage from 
greatly improved observational techniques, facilitating much more 
significant validations of cosmological theories, such as Cold Dark 
Matter (CDM). Although the prime astrophysical importance of the juxtaposition of observations 
and models was assessing the amount of DM
in galaxies \citep{PS:88}, the timed question has become its {\it 
distribution}. This issue directly pertains to the nature of this component, since
a precise knowledge of the density profile $\rho(R)$ is 
likely to be decisive:
the DM nature is likely to determine the halo properties.

Yet, a wealth of observations have not been able to settle but the most 
fundamental issues, such as the validity of the CDM  
scenario on galactic scales (e.g. \citealt{NFW:96}, \citealt{Mo:98}). Some of 
the shortcomings on the observational side were related to the limitations 
of optical rotation curves (internal rms, limited radial extent, dust contamination), 
while radio studies were affected by the well-known limited spatial resolution.
Nevertheless, progress has been made, due to the
increasing number of investigations on dwarf spirals and
low-surface
brightness (LSB) galaxies.
These studies cast serious doubts on one of the fundamental properties
of CDM haloes, namely their 'cuspiness' in the galaxy centres, 
and infer the presence of large core 
radii in the DM density profiles
(see e.g. \citealt{Fl:94}, \citealt{Mo:94}, \citealt{Bur:97}, \citealt{Kr:98}, 
\citealt{Mc:98}, \citealt{SB:00}, \citealt{dB:01}, \citealt{Sa:01}, \citealt{dBB:02}, 
\citealt*{Sa:03}, \citealt*{W:03}, \citealt{Si:03}).
However, due to the many steps in the data analysis, there can be
subtle systematic errors that could distort the results or in any case render
the results very poorly constrained. This has triggered off the recent debate concerning the 
reliability of the data and how well the mass models are really constrained.
There are claims that the observations could actually be
consistent with the dark matter density profiles predicted by
the CDM simulations, not only by considering the {H{\sc i}} data alone 
(\citealt{vdB:00} and \citealt{vdBS:01}), but also by combining
H$\alpha$ and {H{\sc i}} data (\citealt{P:02} and \citealt{S:03}).
This is the reason why particular care should be taken in choosing a suited
sample and in performing the data analysis. Note that recent
simulations (e.g. \citealt{N:03}) do not converge to a well-defined value of the
inner slope down to the resolution limit (about 1 kpc),
even though the slope of the dark matter density profile 
(defined as $ -d{\rm ln}\rho/d{\rm ln}r$) at 1\% of the virial radius 
is still about 1.2 for a typical galaxy.
Notice also that the observational results on
spiral galaxies show a discrepancy with the standard $\Lambda$CDM predictions
well beyond 1 kpc, i.e. well beyond the resolution limit of the simulations.

In this paper we study a {\it sample} of
galaxy rotation curves ideal for deriving the properties of the 
dark matter haloes around galaxies; it consists of five 
late-type bulge-less normal (high surface brightness) spiral galaxies;
compared to the sample of \citet{BS:01} 
we have the same resolution in the central regions, but a larger spatial
extension allowing us to determine the size of the core radii. 
One of the main concerns is that the rotation curve has both a
high spatial resolution
and a large extension, i.e. beyond the optical 
radius; this is typically achieved by combining optical (H$\alpha$) and 
radio ({H{\sc i}}) data (see e.g. \citealt{dBB:02} and \citealt{S:03}).
The former provide the necessary high resolution (1$\arcsec \dots 2 \arcsec$)
while the latter allow us to trace the potential out to large radii,
typically 2--3 times the optical radius. It has been known for a long time
that {H{\sc i}} data usually suffer from the lack of resolution, which can affect
the results, especially in the innermost parts; this problem is nicely reviewed 
by \citet{vdB:00}.

The difficulty in deriving reliable rotation curves from {H{\sc i}} data cubes
resides in the fact that a data cube is 4-dimensional
(RA, dec, flux density and radial velocity), while
the rotation curve is 2-dimensional (rotation velocity vs. radius).
This means
that starting from the data cube we have to take into account two types of considerations: one
concerning the {\it positions} in which the rotation curve should be traced, i.e. how
to reduce the spatial dimensions from (RA, Dec) to galactocentric radius. This is usually achieved by 
taking the position-velocity diagram along the major axis or by fitting concentric 
ellipses around the
centre (tilted-ring modelling of the velocity field). The other 
consideration concerns the {\it rotation velocities} that should be attributed to these galactocentric radii: 
at the positions defined above we want to associate only one of the possible velocities
that could in principle be derived from the spectra at single points (flux density vs. radial velocity),
i.e. the radial component of the rotation velocity at that radius. 
This is usually done via different methods,
most of them assuming symmetry of the profiles, like the first-moment analysis 
(the intensity weighted mean), or the single Gaussian fitting; therefore, when the profiles
are not symmetric, i.e. in galaxies with a high inclination and/or a poor to intermediate resolution,
these methods cannot be applied.

\begin{table*}
\caption{Physical parameters of the galaxies. M$_I$ is the I-band absolute 
magnitude and M$_{\rm HI}$  is the total {H{\sc i}} mass, 
$r_{\rm HI}$ is the {H{\sc i}} radius (defined as the radius 
at which the surface density drops below 1 M$_{\odot}$ pc$^{-2}$) and $r_{d}$
is the exponential scale length. $r_f$ is the farthermost radius with data and
$r_{opt}$ is the optical radius (defined as 3.2$r_d$) and RR us the radio resolution scale. 
The dynamical mass M$_{dyn}$ is determined at $r_{f}$.
\vspace{0.3cm}}
\begin{tabular}{l l l l l l}
\hline
\hline
Galaxy name                      & ESO 116-G12        & ESO 287-G13       & ESO 79-G14 & NGC 1090 & NGC 7339    \\ 
\hline 
M$_I$ (mag)                      & -20.0              & -21.7             & -21.4      & -21.8    & -20.6       \\
M$_{\rm HI}$  (M$_{\odot}$)      &1.5$\times 10^{9}$  & 1.1$\times 10^{10}$&3.5 $\times 10^{9}$ &8.5$\times 10^{9}$ &5.5$\times 10^{8}$    \\
Inclination ($^{\circ}$)         & 74                 &78                 &84          &64        &79           \\
$r_{\rm HI}/r_{d}$          & 6.7                &7.4                & 4.8        &8.8       &4.1          \\
$r_f$ (kpc)                 & 11.4               & 24.9             & 17.7       &     29.7   & 5.4        \\
$r_{opt}$ (kpc)             & 5.4               & 10.5            & 12.4        &     10.9   & 4.9         \\
$r_f$/RR                    & 5.4               & 5.8            & 5.0        & 10.6      &  4.6          \\  
M$_{dyn}$ (M$_{\odot}$)     & 3.3$\times 10^{10}$ & 1.9 $\times 10^{11}$ & 1.3$\times 10^{11}$ & 1.8$\times 10^{11}$ & 3.1$\times 10^{10}$ \\  
\hline
\label{physical}
\end{tabular}
\end{table*}

We have developed a new method for both of these considerations:
a) concerning the positions, we traced the rotation curve along the ridge of the warp 
(this step is discussed in \citealt{V:04}),
when the velocity field was not sufficiently sampled (i.e., for the galaxies ESO 116-G12,
ESO 287-G13, ESO 79-G14 and NGC 7339) and the
tilted-ring modelling of the velocity field was therefore not possible;
b) for the velocities 
we only considered the velocity side of the profiles opposite to the systemic velocity, and then
corrected for the effects that artificially broaden the profile, i.e.:
the turbulence of the ISM, the instrumental velocity resolution and
the beam-broadening. 
The present paper concentrates on step b).

When both steps a) and b) were applied we called our method WAMET (WArped Modified Envelope
Tracing method); when only step b) was applied and a velocity field was constructed (i.e., when
the velocity field was sufficiently sampled, the case of NGC 1090) we called our method MET (Modified
Envelope Tracing method). These approaches were tested by
constructing artificial data cubes based on geometrical models of the {H{\sc i}}
disc that were iteratively compared to the observed data cubes.  
The MET/WAMET
method proves to provide a better initial estimate of the rotation curve then
more traditional methods, such as the first-moment analysis and
the single-Gaussian fitting. Modelling the data cubes results in a powerful way
to test and improve the derived rotation curves, as shown e.g. by \cite{Gea:03}.

This work is structured as follows: in Section 2 we present the sample and optical data that 
were used in this study, in Section 3 we describe the {H{\sc i}} observations 
and the data reduction and Section 4 shows the data analysis; in Section 5 
we present the results of data analysis, the new method for extracting the
rotation curves is shown in Section 6 and the comparison between the observed
and the modelled data cubes is presented in Section 7. 
In Section 8 we introduce the different
mass models that were considered; the results
of the mass decompositions are shown in Section 9 and in Section 10 
we draw the conclusions concerning the derivation of the rotation curves and 
the dark matter haloes around
the galaxies of our sample.

Throughout this paper we adopted a value for the Hubble parameter $H_0$=75 km s$^{-1}$ Mpc$^{-1}$.

\begin{table*}
\centering
\begin{footnotesize}
\centering
\caption[]{
Optical and radio resolution scales (indicated 
respectively as OR and RR), and $r_{f}$, in units of the scale radii
of the Burkert and NFW density profiles. 
\vspace{0.3cm}
\label{rs}}
\begin{tabular}{l c c c c c c}
\hline
\hline
Galaxy       &OR/$r_{core}$     & RR/$r_{core}$ &OR/$r_{s}$  &RR/$r_{s}$   &   $r_{f}$/$r_{core}$  & $r_{f}$/$r_s$     \\
\hline
ESO\,116-G12 & 0.03              & 0.49  & 0.02   & 0.29         & 2.7        & 1.6                    \\
ESO\,287-G13 & 0.01              & 0.18  & 0.02   & 0.29         & 1.0        & 1.7                    \\
ESO\,79-G14  & 0.04              & 0.43  & 0.02   & 0.24         & 2.1        & 1.2                    \\
NGC\,1090    & 0.04              & 0.32  & 0.03   & 0.23         & 3.4        & 2.5                    \\
NGC\,7339    & 0.02              & 0.15  & 0.01   & 0.08         & 0.6        & 0.3                    \\
\hline
\end{tabular}
\end{footnotesize}
\end{table*}

\section{The sample}

\subsection{The relevance of the present sample}

The present sample is ideal for the study of dark matter within galaxies,
not only for the characteristics of the galaxies, but also for the quality
of the data. Table~\ref
{physical} presents the galaxy sample and its 
physical characteristics. 
This sample allows us to investigate 
a range of luminosities approximately 1.5 mag fainter than L$_*$, 
ideal for investigating the dark matter component, because it starts dominating the
kinematics at relatively small radii.

In Table \ref{rs}, the crucial characteristics of the kinematical data are shown:
optical data allow us to sample very well the region inside the characteristic
radii of various dark matter distributions, 
while {H{\sc i}} data allow us to probe a region {\it larger} than these radii, which permits us to
investigate the core radius phenomenon at large distances and consequently to
constrain the size of the core radii (note, however, the case of NGC 7339 discussed in Appendix A).

The rotation curves used here consist of the H$\alpha$ data for the inner
parts and the {H{\sc i}} data for the outer parts. The claim
of unreliability of observed kinematics put forward by \citet*{Rh:03}  
does not apply here: 
a) the derivation of the {H{\sc i}}
rotation curve was performed in a more thorough and reliable way than the standard
tilted-ring analysis on the velocity field; b) the galaxies studied
here were at an inclination optimal to
minimise projection effects; 
c) in the regions were they coexist, the H$\alpha$ and
{H{\sc i}} data (emerging from different physical processes) do
agree, implying a high quality of the data\footnote{Note, however, that this is
not entirely true for NGC 1090, in which for some radii the H$\alpha$
lie $\sim$10 km s$^{-1}$ above the {H{\sc i}} data.}.
The small internal scatter inside the each radial bin
indicates that, for the galaxies of our sample, the effects of non-circular motions
are negligible (even though a more thorough investigation should involve
2-dimensional H$\alpha$ data), differently from 
the case shown by \citet{Sw:03}. 
Moreover, it can be shown (see e.g. \citealt{dB:03}) that 
the bias towards lower velocity gradients in the rotation curve due to
a misalignment between the slit with the kinematical axis
is likely to be small.
\subsection{Optical data}
The five galaxies of this sample were chosen from the 
sample of 967 galaxies with optical rotation curves presented by
\citet{PS:95}, according to the following criteria:
a) they belong to the ``excellent'' subsample (i.e. the
two sides are symmetric, the data are extended out to at least 
$r_{opt}$ and the number of data points is $\geq$ 30);
b) they have a reasonably high total {H{\sc i}} flux; c) they
have a relatively low I-band luminosity ($M_{\rm I} > -21.8$),
d) they have a large angular size, and e) they have an inclination $i$
suitable for H{\small I} studies ($50^{\circ} < i < 85 ^{\circ}$).
For each galaxy, the raw H$\alpha$ data were binned in
groups of 4 to 6 \citep{PS:95}, and the given error is 
the uncertainty on the average value inside the radial bin.
We took a ``minimum error'' equal to half the average error
in order to avoid data points with unrealistically low
errors that could bias the rotation curves decompositions;
for the same reason, when a data point was clearly
inconsistent with the two neighbouring points and the general
trend of the rotation curve, the value of its error was increased.
The complete description of the optical observations
(spectroscopic and photometric)
can be found in \citet*{Ma:92} and \citet{PS:95}.

\section{{H{\sc i}} Observations and reduction}
\label{sectobs}

\begin{table*}
\centering
\begin{footnotesize}
\centering
\caption{{H{\sc i}} Observational parameters}
\vspace{0.3cm}
\label{observational}
\begin{tabular}{l l l l l l}
\hline
\hline
Galaxy name  & ESO 116-G12                             & ESO 287-G13                              & ESO 79-G14           & NGC 1090       & NGC 7339 \\ 
\hline
RA           & 3$^{\rm h}$13$^{\rm m}$03$^{\rm s}$.9  & 21$^{\rm h}$23$^{\rm m}$13$^{\rm s}$.7  & 1$^{\rm h}$02$^{\rm m}$51$^{\rm s}$.2& 2$^{\rm h}$46$^{\rm m}$33$^{\rm s}$.9           &    22$^{\rm h}$37$^{\rm m}$46$^{\rm s}$.9  \\
Dec          &$-57\degr 23\arcmin27\arcsec$             & $-45\degr46\arcmin20\arcsec$               &$-65\degr36\arcmin36\arcsec$&$ -0\degr14\arcmin50\arcsec$         & $23\degr47\arcmin12\arcsec$ \\
Number of channels  & 256                              & 256                                      & 256                  & 2 $\times$ 64           &  2 $\times$ 64\\
Channel separation (kHz) & 31.2                        & 31.2                                     &31.2                  &24.4              & 24.4 \\
Synth. beam (arcsec) & 30.4 $\times$ 24.4              &32.6 $\times$ 24.6                        &27.4 $\times$  23.0   &17.6 $\times$ 14.7 & 13.9 $\times$ 13.5\\
Synth. beam (kpc)    & 2.3 $\times$ 1.8               &5.6 $\times$ 4.2                         & 4.0 $\times$ 3.4      & 3.1 $\times$ 2.6 & 1.2 $\times$ 1.2 \\ 
Noise (mJy beam$^{-1}$) & 1.2                       &1.2                                       & 1.1                 & 0.8                &0.7  \\
\hline
\end{tabular}
\end{footnotesize}
\end{table*}

We observed ESO 116-G12, ESO 287-G13 and ESO 79-G14 with
the Australia Telescope Compact Array in the 750 m and 1.5 km
configurations; the resulting baselines range from 31 m to 1500 m.
The galaxies were observed for 12 hrs in each configuration.
The correlator setup yielded
512 channels of width 15.6 kHz, resulting in a total bandwidth of 8 MHz.
The data were subsequently Hanning-smoothed to have a better 
signal-to-noise ratio. 
The final spectral resolution was 31.2 kHz, corresponding to a velocity 
resolution of 6.6 km s$^{-1}$.

NGC 1090 and NGC 7339 were observed with
the Very Large Array in the C configuration, with 
baselines ranging from 35 m to 3400 m.
The galaxies were observed for 8 hrs 
with a correlator mode which had 4 IFs, each IF pair
having slightly different frequency centres; this was done in order
to find a compromise between velocity resolution and bandwidth.
We had $2 \times 64$ channels of width 24.4 kHz
and a total bandwidth of 2.685 MHz; taking into account the 
overlapping channels, the total number of channels is 112. 
The resulting velocity resolution 
is $\sim 5.2$ km s$^{-1}$.

The standard flagging and calibration of the $u-v$ data from
the ATCA observations
was performed within the software
package MIRIAD (\citealt{S:95}), while the 
the VLA data observations were flagged and calibrated with the 
Astronomical Image Processing Software (AIPS).
The rest of the reduction was performed in MIRIAD.  

Using a number of line-free channels (10 \dots 40) on 
each side of the line, the continuum contribution to the $u-v$ data
was calculated performing a linear fit; this contribution
was then subtracted from the data. Then, 
we produced a data cube with ROBUST parameters \citep{Br:95} ranging 
from 0 to 0.3,
to reach an ideal compromise between high signal-to-noise ratio and 
small beam size;
these values correspond to a weighting scheme of the $u-v$ data
intermediate between uniform and natural weighting. 
The data were then CLEANed, first on a small region with emission
and then on the whole map, with a total of 2500 iterations per channel map.
The data were then restored with a Gaussian beam.
Depending on the galaxies, we obtained beam sizes ranging
from $13 \arcsec$ to $33\arcsec$. The observational parameters are listed
in Table \ref{observational}.

\begin{figure*}
 \centerline{\psfig{figure=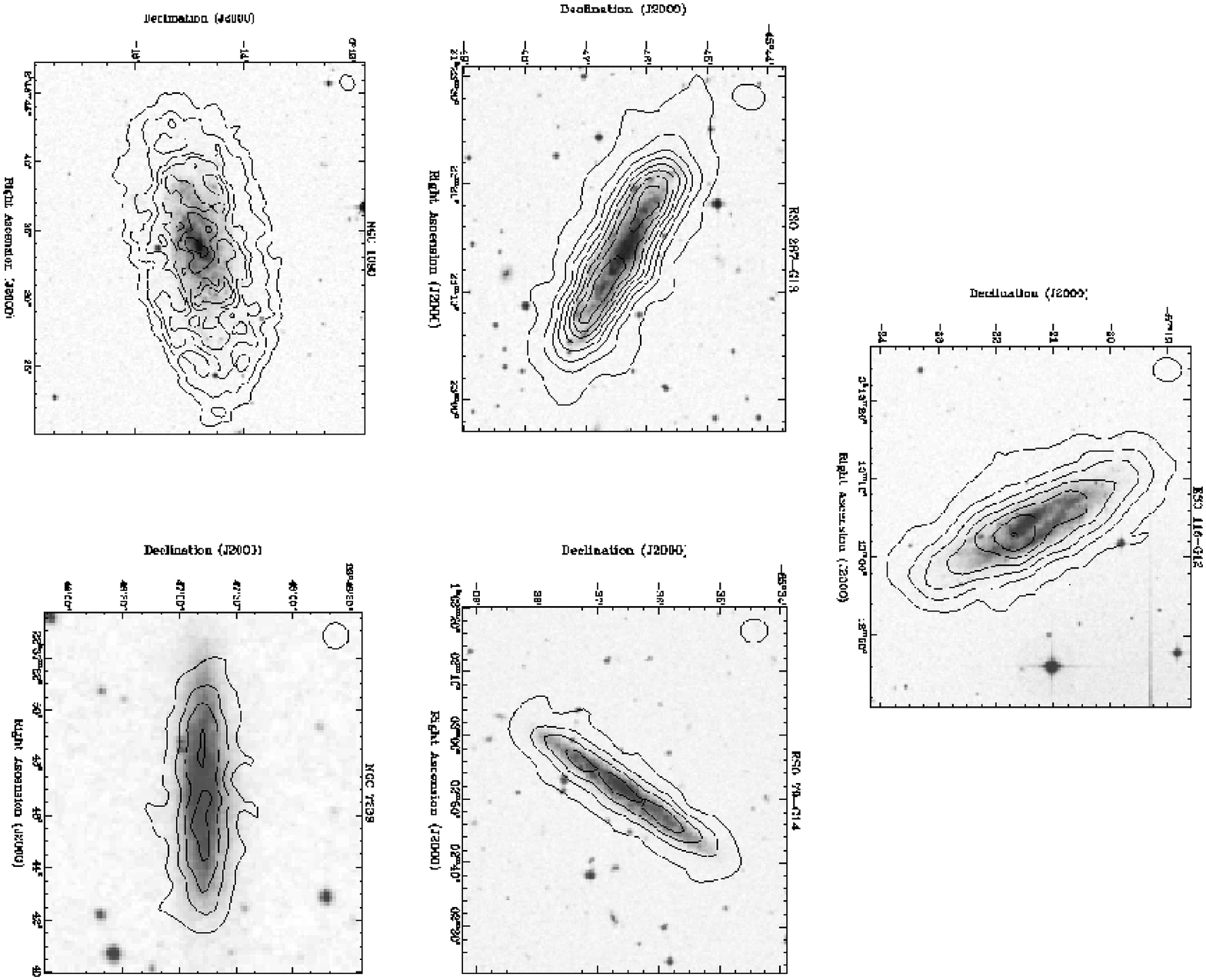,width=17.35cm,angle=90}}
  \caption{Optical DSS images (greyscale) superimposed with the
  H{\sc i} total intensity map (contours). The first contour is
  the ``pseudo 3-$\sigma$'' defined similarly to \citet{V&S:01} and 
  is equal to: $1.0 \times 10^{20} {\rm atom~ cm}^{-2}$ for ESO 116-G12; 
$1.1 \times 10^{20} {\rm atom~ cm}^{-2}$ for ESO 287-G13;
$1.7 \times 10^{20} {\rm atom~ cm}^{-2}$ for ESO 79-G14;
$1.5 \times 10^{20} {\rm atom~ cm}^{-2}$ for NGC 1090;
$3.4 \times 10^{20} {\rm atom~ cm}^{-2}$ for NGC 7339;
the remaining contours are (15, 30, 45,...)~$\times~ \sigma$, except for
NGC 1090 and NGC 7339 where they are (6, 12, 18,...)~$\times~ \sigma$.
The beam is shown in the upper left corner.
}
\label{radopt2}
\end{figure*}

\begin{figure}
  \centerline{\psfig{figure=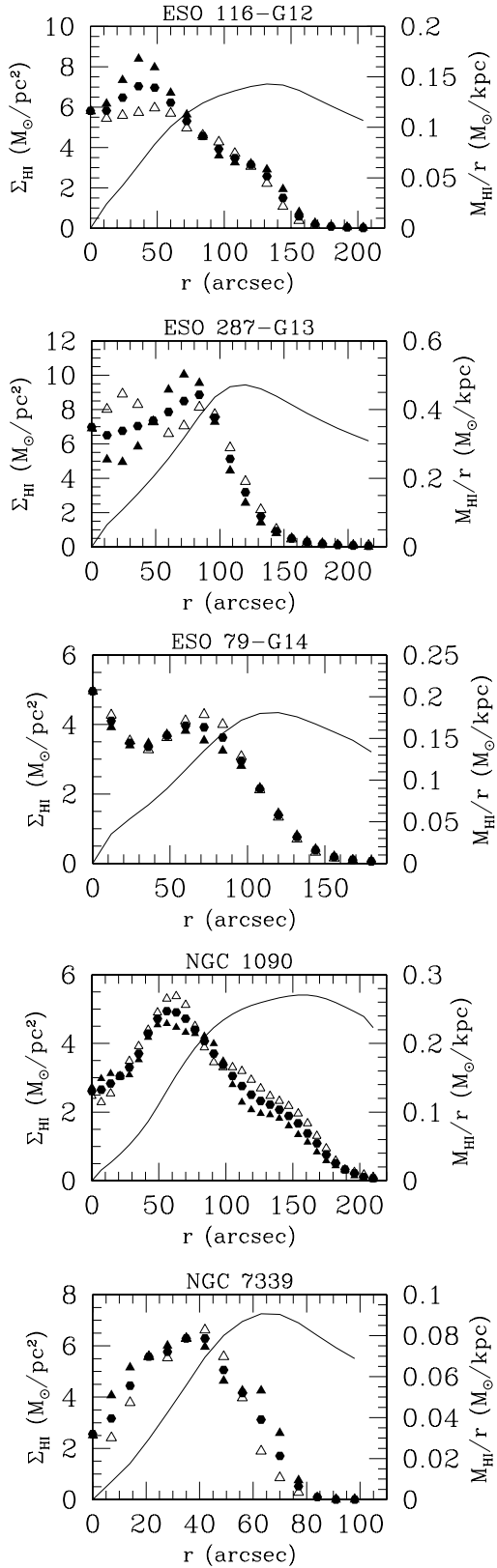,width=6.85cm}}
  \caption{Radial distribution of the neutral hydrogen surface density
($\Sigma_{\rm HI}$):  
the filled triangles denote the approaching side,
the empty triangles the receding side and the filled circles 
the average. The solid line refers to the ratio 
${\rm M}_{\rm HI}(<r)/r$, where ${\rm M}_{\rm HI}(<r)$ is the H{\sc i} mass inside
radius $r$.
 }
\label{raddens}
\end{figure}

\begin{figure*}
  \centerline{\psfig{figure=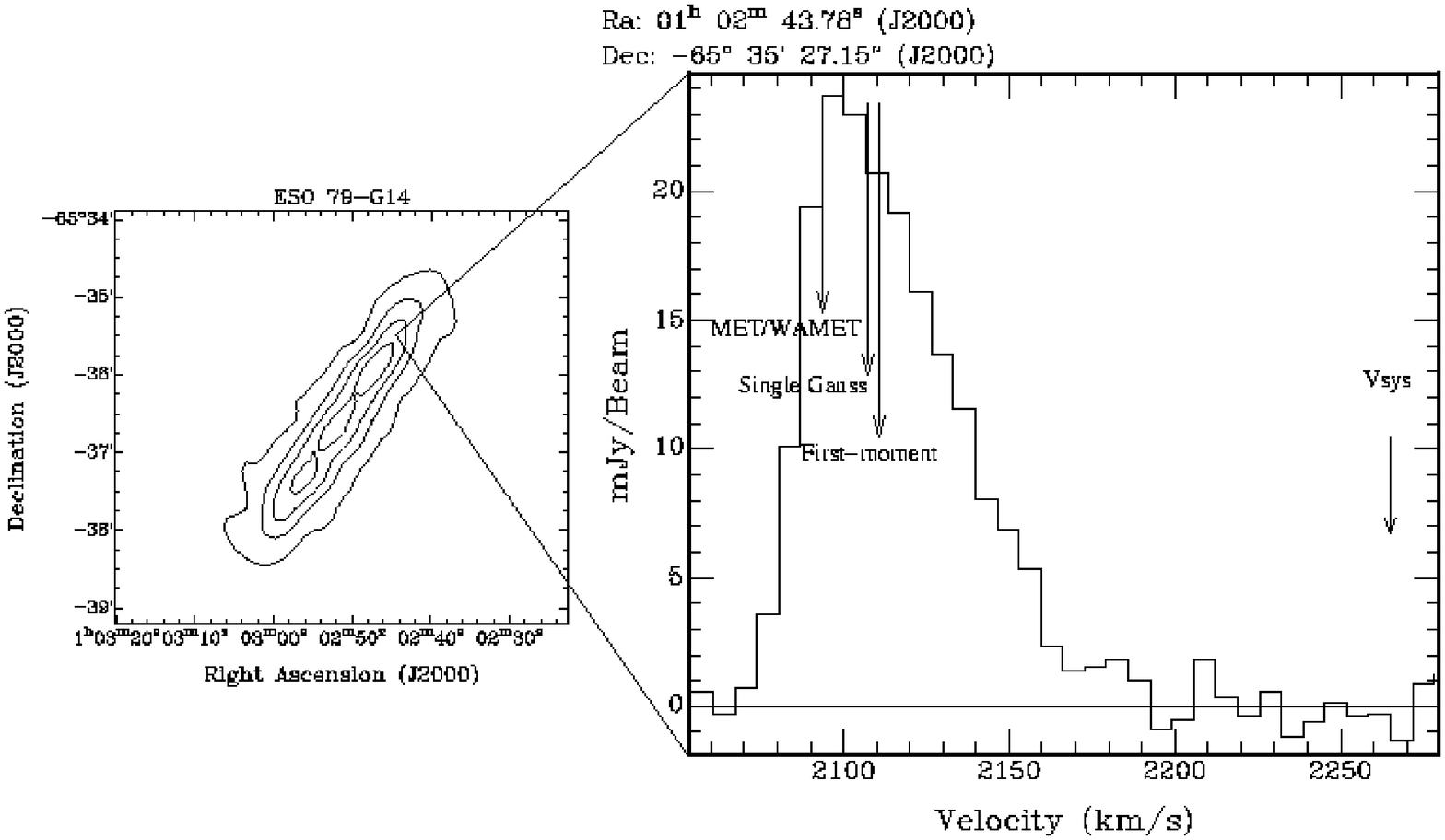,width=17.7cm}}
\caption{A typical spectrum of the galaxy ESO 79-G14 
at an intermediate galactocentric distance.
The arrows indicate the systemic velocity as well as the positions of the 
velocities derived from the MET/WAMET method, the first-moment
analysis and by fitting
a single Gaussian to the profiles.}
\label{spectrum}
\end{figure*}

\section{Data analysis}

From the CLEANed cubes, new cubes with lower resolution were constructed
(with a beam size about twice the original). This was done in order to distinguish 
genuine emission from spurious one; we considered as having reliable emission all the regions with an
intensity $> 2 \times \sigma$ in the low-resolution cube;
spurious noise was also removed. 
Moment-0 maps (total intensity maps) were derived based on the ``masked'' cubes. They are shown in
Fig.~\ref{radopt2}, superimposed with optical DSS images. The first contour
is the ``pseudo 3-$\sigma$'' level defined in a way similar to \citet{V&S:01}:
we define the noise at a certain position in
the total intensity map as:

\begin{equation}
\sigma_N^h=\left(\frac{N}{2}-\frac{1}{8}\right)^{\frac{1}{2}}
\frac{4}{\sqrt{6}} \sigma^h
\end{equation}

where $N$ is the number of channels that have been added at each position 
in the total intensity map and $\sigma^h$ is the noise in the 
Hanning-smoothed channel maps. The ``pseudo 3-$\sigma$'' level is
obtained by averaging all the pixels with a signal-to-noise ratio
between 2.75 and 3.25. 

\section{Observational results}

From Fig.~\ref{radopt2} we note that the galaxies 
have a relatively high inclination
and the {H{\sc i}} emission has a large (up to $\sim$ 3 times) extent compared to the
optical disc. 
The {H{\sc i}} masses ($M_{\rm HI}$) given in Table \ref{physical} were
calculated from the total intensity maps after primary-beam 
correction using the following relation:

\begin{equation}
M_{\rm HI}=2.36 \times 10^5 D^2 \int S {\rm d} V
\end{equation}

where $M_{\rm HI}$ is in ${\rm M}_{\odot}$, $D$ is the distance of the galaxy in Mpc
and $\int S {\rm d} V$ is the total {H{\sc i}}
flux density in the total intensity map, measured in Jy km s$^{-1}$.
The {H{\sc i}} masses are in agreement
with previous single-dish measurements to within 20\%, implying that we are not
losing much extended emission due to missing zero spacings.

The radial distribution of the neutral hydrogen surface density
is computed by integrating the total intensity map over concentric ellipses 
for NGC 1090 and 
by means of the Lucy method \citep{W:88}
for the other galaxies. The results are shown in Fig.~\ref{raddens}, where
one can notice that the data reach the region where ${\rm M}_{\rm HI}(<r)/r$
starts to decrease.

Since the velocity field of NGC 1090 is sufficiently sampled,
it could be subjected to tilted-ring modelling \citep{B:89}. 
In this case the MET method was applied
(see the next Section) to construct the velocity field and then the first estimate of the 
rotation curve was derived by performing a tilted-ring fit to the velocity
field.
In the other cases the combination of high inclination
and a large beam (compared to the size of the galaxy)
requires the use of the WAMET method:
we traced the rotation curve 
along the warp on points defined like in \citet{G:02} and
implemented for studies of the kinematics by \citet{V:04} on the total 
intensity map,
we fit Gaussians to the density profiles parallel the minor axis,
defining the position of the peak of the Gaussian as the ``ridge'' of the warp. 

\begin{figure}
  \centerline{\psfig{figure=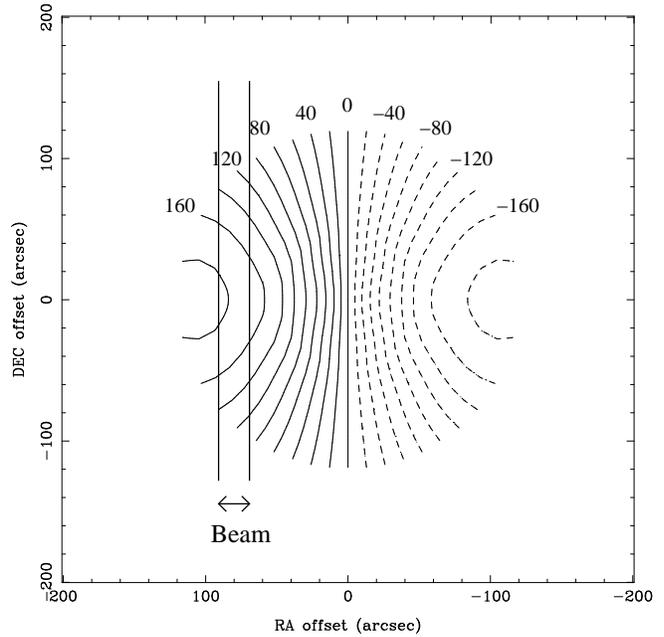,width=8.85cm}}
\caption{Model of the radial velocities in the plane of the galaxy
ESO 79-G14 and the regions that would be intercepted by the beam
in the edge-on case.}
\label{plane}
\end{figure}

Concerning the analysis of the velocity profiles, particular
care was taken in deriving the rotation velocity because
they were in general asymmetric.
In Fig.~\ref {spectrum} we show an example of a 
typical profile at an intermediate galactocentric distance: it
is not symmetric and it has a tail towards the systemic velocity. This
is what we expect in case of a highly inclined galaxy,
because (see Fig.~\ref {plane}) what we observe is the integration along 
a large portion of the disc, so that material with lower radial velocities
also contributes to the velocity profile. Note, however, that Fig.~\ref {plane}
represents the extreme edge-on case: the galaxies studied
here were chosen to have lower inclinations, so the portion of 
the disc intercepted by the line-of-sight is smaller.

Moreover, from recent results (\citealt*{S:97}, \citealt{F:01}),
it seems that at least in some spiral galaxies there is evidence for  
neutral hydrogen at several kpc away from the galactic plane that rotates more
slowly than the gas in the disc; this effect also produces an asymmetry of 
the velocity profiles 
in galaxies that do not have a high inclination.
The result is that the standard methods (e.g. first-moment analysis 
and single-Gaussian fitting), which implicitly
assume symmetry of the profiles, provide velocities that 
are biased towards the systemic velocity (see Fig.~\ref{spectrum})
and cannot be used to determine the true rotation velocity
in the galaxies of our sample.

\section{The Modified Envelope-Tracing method}
\label{wamet}

In order to solve the problems discussed in the previous Section,
we built a method similar to the Envelope-Tracing
method (e.g. \citealt{S:79}, \citealt{So:97});
the aim is to fit only the side of the profiles which we are
interested in, i.e. the extreme velocity side (with the assumption
that the gas is present at these extreme velocities).

The first step was to fit a half-Gaussian from the peak
of the profiles to their extreme velocity
side (i.e. the side opposite to the
systemic velocity), considering the velocity at half maximum, the
terminal velocity ($V_{t}$).
The rotation velocity $V_{rot}$ at a certain position is then given by:
\begin{equation}
V_{rot}=\frac{|V_{t}-V_{sys}|-0.5\sqrt{(\delta V_{ISM})^2 + 
(\delta V_{obs})^2 + (\delta V_{b})^2}}{sin~i}
\end{equation}

where $i$ is the inclination, $V_{sys}$ is the systemic velocity, 
and the terms indicated with $\delta V$
describe the profile broadenings
(in terms of full width to half maximum, FWHM):

$\delta V_{ISM}$
is the broadening due to the turbulence of the 
interstellar medium, by assuming a constant velocity dispersion of the ISM, 
$\sigma_{ISM}=10 $ km s$^{-1}$ (a typical value for spiral
galaxies), which implies:
 
\begin{equation}
\delta V_{ISM}=\sigma_{ISM}\sqrt{8{\rm ln}2}= 23.5 ~{\rm km~ s}^{-1}. 
\end{equation}

Notice that following \citet{K:93} (i.e. taking $\sigma_{ISM}$ going
from 12 to 7 km s$^{-1}$)
would not significantly affect the derived rotation curves.

$\delta V_{obs}$ is the instrumental contribution
to the profile broadening, 
which we set equal to the channel resolution: 6.6 km s$^{-1}$ 
for the galaxies observed with the ATCA and 5.2 km s$^{-1}$ 
for those observed with the VLA.

$\delta V_{b}$ is an  
estimate of the broadening of the profiles
due to the beam: as is evident in Fig.~\ref{plane}, 
a larger beam will sample a larger portion
of the velocity field and will thus broaden the profiles,
even on their extreme velocity side.
According to \citet{S:79}, a lower limit to $\delta V_{b}$
can be given 
by setting it to zero, yielding
an upper limit to the rotation velocity.
To estimate an upper limit to $\delta V_{b}$, 
we assume as in \citet{B:97}, 
that the expected beam uncertainty
in the profile widths
is $2\cdot[V(r \pm  \theta_{b}/2)-V(r)]$, 
where the ``$+$'' 
corresponds to a positive gradient of the velocity field and
``$-$'' to a negative one; $\theta_{b}$ is the 
beam FWHM.
 
With an upper limit
and a lower limit 
to $\delta V_{b}$, we decided to 
estimate $\delta V_{b}$ using their average, i.e:

\begin{equation}
\delta V_{b}=V(r \pm \theta_{b}/2)-V(r)
\end{equation}

MET/WAMET has the following advantages:

\begin{itemize}

\item[$\bullet$]{The profiles are asymmetric, thus methods assuming symmetry cannot be 
used to derive the rotation velocity; they would underestimate the rotation velocities, 
especially in the inner parts.}
\item[$\bullet$]{We are only interested in the extreme velocity side of the
profiles because it is not (or little) affected by projection effects.}
\item[$\bullet$]{It also estimates the correction for the beam broadening
of the profiles, which can be substantial in regions 
where the gradient of the velocity field is high.}
\end{itemize}

The implemented method (WAMET) for warped galaxies can be applied to 
galaxies where the sampling of the
velocity field is poor, and it enables us to trace the rotation curve along any
possible warp instead of keeping a fixed position angle, like
in methods based on the analysis of the position-velocity diagram; 
the shortcomings of keeping a fixed position angle are discussed in \citet{V:04}.

\section{Rotation curves and models}

\begin{figure*}
  \centerline{\psfig{figure=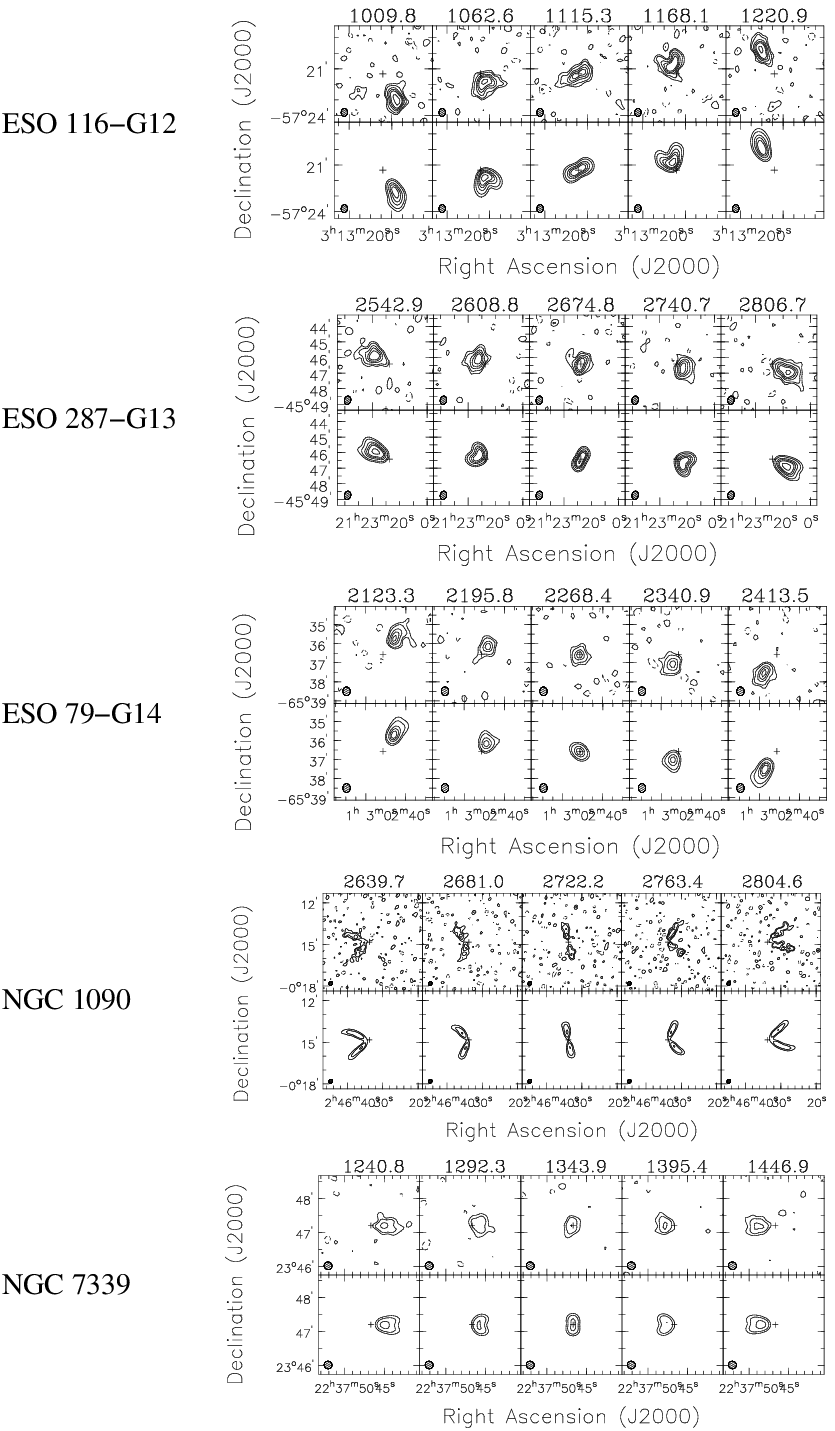,width=12.7cm}}
  \caption{
Five representative observed channel maps (upper panels) with the correspondent
channel maps of the model data cubes (lower ones). The heliocentric radial velocities are
indicated above each plot. The central map has a velocity closest 
to systemic. The cross indicates the centre of the galaxy.
Contours are $-4\sigma, -2\sigma, 2\sigma, 4\sigma$, then 10, 15, 20, 30,
50 mJy beam$^{-1}$ for the ATCA galaxies, and $-4\sigma, -2\sigma, 2\sigma, 4\sigma$,
then 6, 10, 15, 20, 30, 50 mJy beam$^{-1}$ for the VLA galaxies. 
The beam is shown in the lower left corner.
}
\label{channels2}
\end{figure*}

\begin{figure*}
  \centerline{\psfig{figure=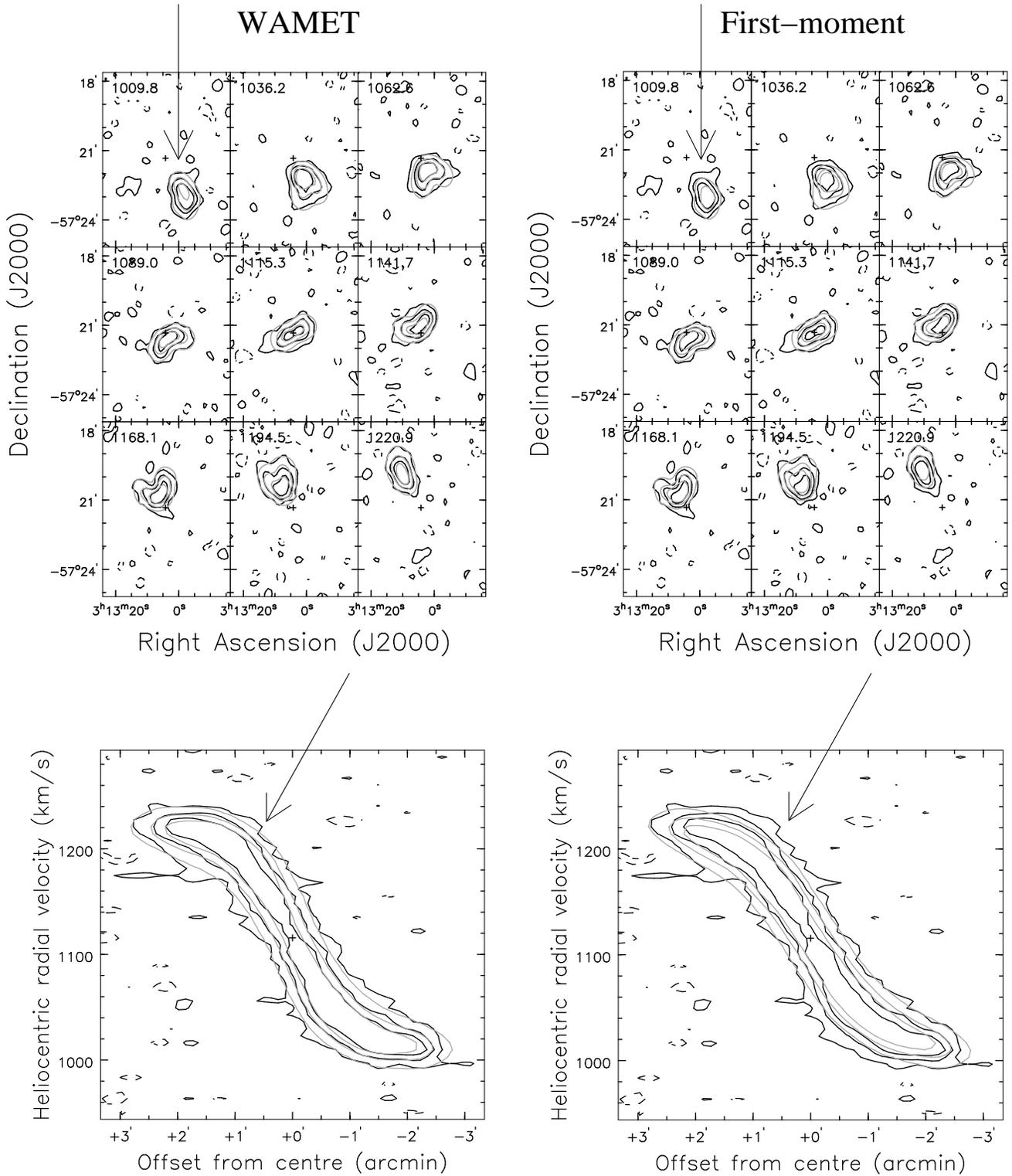,width=17.6cm}}
  \caption{
Comparison between the {\it observed} data cube and two {\it model} data cubes
for ESO 116-G12: 
the MET/WAMET rotation curve (left) and the 
first-moment analysis rotation curve (right);
the black contours trace the observed data cube and the
grey contours trace the model data cubes. 
On the top we show some representative channel maps 
(the heliocentric radial velocities are shown in the upper left corners) and
at the bottom the position-velocity diagrams along the major axis. 
The contours are -2.5, 2.5, 10, 20, 50 mJy beam$^{-1}$.
The cross indicates the centre of the galaxy
and the arrows show some key positions where the difference 
between the two model data cubes is noticeable.}
\label{compar}
\end{figure*}

\begin{figure*}
  \centerline{\psfig{figure=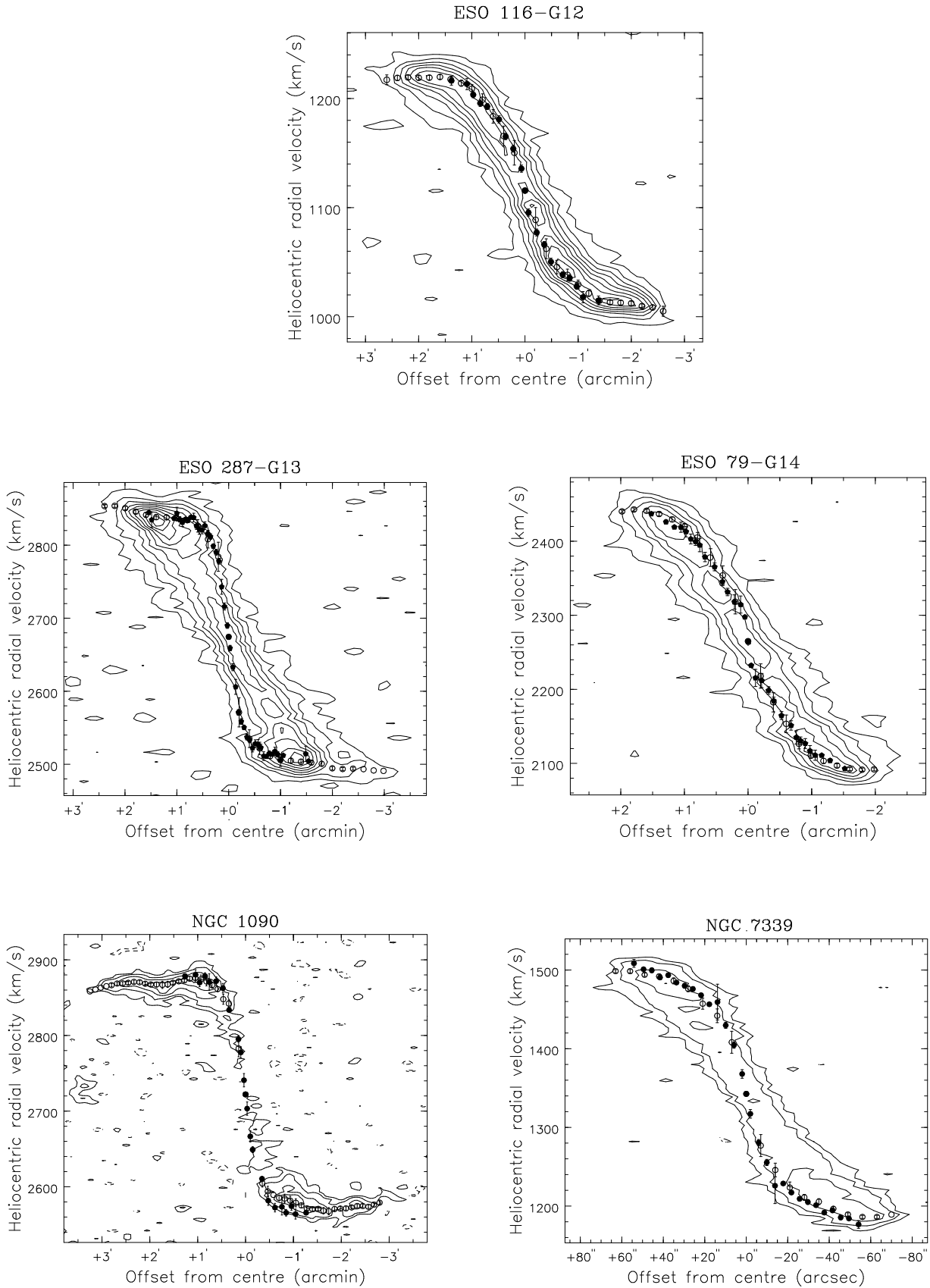,width=16.3cm}}
  \caption{H$\alpha$ (filled circles) and H{\sc i} (empty circles)
rotation curves projected onto the H{\sc i} position-velocity diagram
along the major axis.
The contours are n $\times~ \sigma$, with n=2,4,6,... 
 }
\label{pvrc2}
\end{figure*}

\begin{figure*}
  \centerline{\psfig{figure=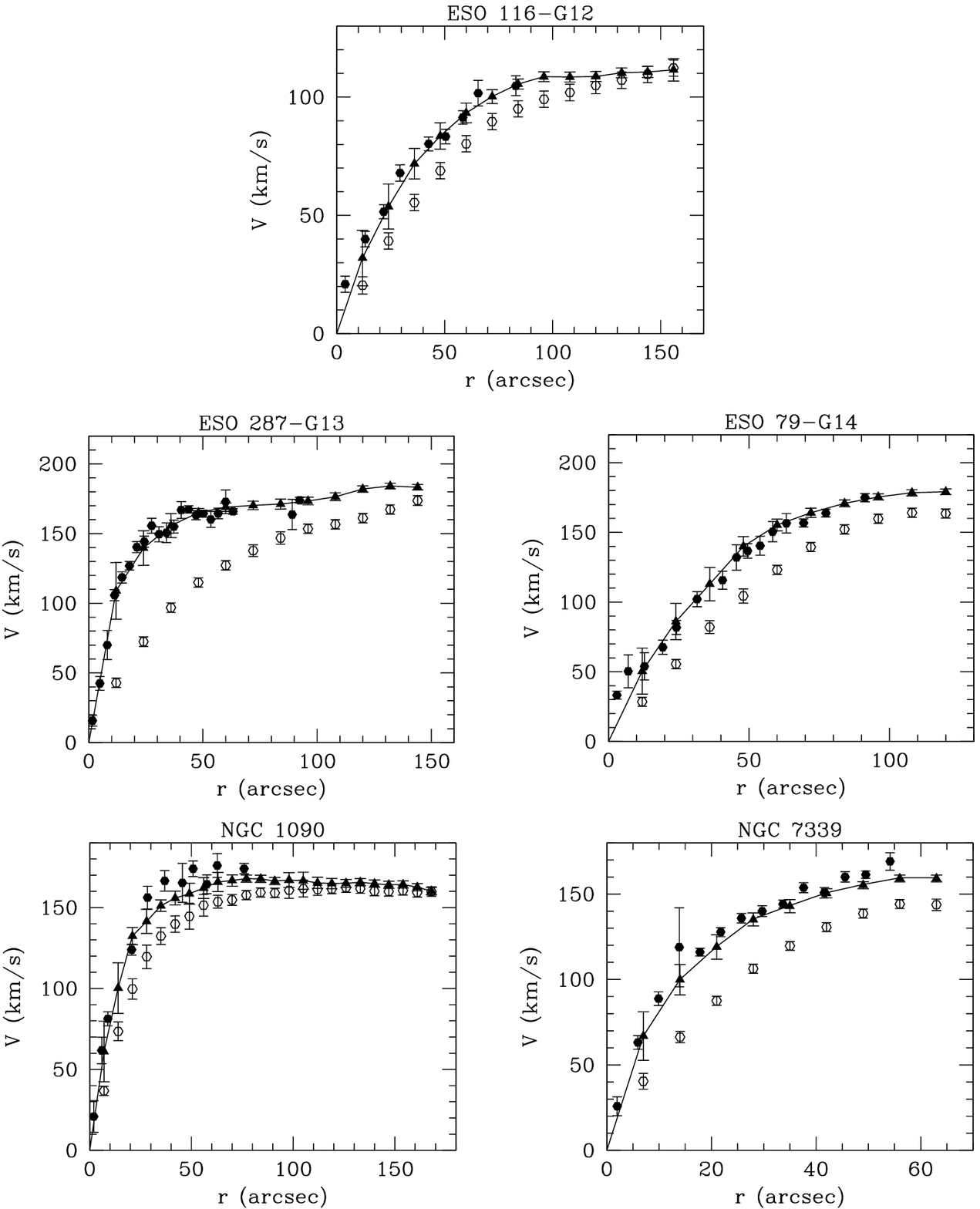,width=17.7cm}}
  \caption{The rotation curves: the H$\alpha$ (filled circles)
rotation curves, the {H{\sc i}}  
corrected MET/WAMET rotation curves (filled triangles) and
the {H{\sc i}} first-moment analysis rotation curves (empty circles).
The line connects the points of the corrected MET/WAMET rotation curves.
 }
\label{rc}
\end{figure*}

In order to derive the rotation curves, we applied the tilted-ring 
modelling of the velocity field for NGC 1090. For the other galaxies we determined the kinematical 
centre and the systemic velocity by minimising the differences between the two
sides. The errors are the maximum of the three following values: the difference
between the velocities of the approaching and the receding sides, our correction for 
beam broadening of the profiles, and a ``minimum error'' 
equal to ($2/sin~i$) km s$^{-1}$.

The MET/WAMET rotation curve served as an input to 
construct model data cubes with the task GALMOD within GIPSY \citep{vdH:92}; 
these models
have been then compared to the observed data cubes. 
Such models are built with the assumption that the neutral hydrogen moves along
circular orbits; a series of geometrical and physical parameters, among which
the rotation curve, allows us to create model observations of the {H{\sc i}}
discs. These models, when compared to the real data cubes,
verify whether they are a fair representation of the {H{\sc i}} disc.
Some input parameters are inferred directly from the observations and
kept fixed (e.g., the
surface density profile), while others (e.g., the kinematical 
centre and the systemic velocity) are first estimated from the
observations and then adjusted in order to optimise the match between
the observed and the model cubes.

In two cases (ESO 287-G13 and NGC 1090, the galaxies with the steepest rotation curves) the 
rotation curve derived with MET/WAMET had to be slightly modified in order to better reproduce
the observed data cubes. We found out that it was necessary to increase the value of the 
rotation velocity of the 2--3 innermost 
points by values of the order of their $1~ \sigma$ error; this is probably due to the fact that,
even with the MET/WAMET method, when the rotation curve rises steeply we are not able
to retrieve precisely the true rotation velocity; this is however possible by modelling the data cube. 
In Fig.~\ref{channels2},
we display some channel maps (observed and modelled) to show that
with our final rotation curves we are able to provide a very good reproduction of the 
observations. 
All the galaxies of our sample were successfully modelled with a small warp in
the outer parts or with no
warp at all. In particular, we found necessary to include in our model parameters a
5$^{\circ}$ clockwise warp in ESO 116-G12, a 2$^{\circ}$ clockwise warp in
ESO 79-G14 and a 2$^{\circ}$ anti-clockwise warp in NGC 1090. Concerning the inclination
as a function of radius, it was kept constant at the values given in Table \ref{physical},
except for the galaxy ESO 79-G14, where a 5$^{\circ}$ decrease of the inclination in the
outer parts was necessary to reproduce the shape of the extreme channels. The errors
on the derived orientation angles are of a few degrees (\citealt{Gea:03}).

Using the rotation curves derived from other methods provided data cubes
that were clearly different from those observed.  In Fig.~\ref{compar},
for the galaxy ESO 116-G12,
we show some channel maps and the position-velocity diagram along
the major axis of the observed data cube, of a model created from the corrected MET/WAMET
rotation curve and of a model created from the first-moment rotation curve. The latter
clearly fails in reproducing some of the observed features, like the shape of 
the extreme channels and the slope of the position-velocity diagram. Fitting 
single Gaussians to the velocity profiles gives very similar results to the 
first-moment rotation curve.

In order to assess the validity of a model we used a number of tools to visualise
the different properties of the data cube. In particular, the model
has to simultaneously match the intensity and shape of the emission in the
single channel maps, the column density distribution in the moment-0 map,
and the intensity and shape of the emission in the position-velocity diagrams,
along the major and minor axis and on slices displaced with respect to the 
galactic centre.
The comparison
of model data cubes to the observations is a very powerful and crucial test to check 
quality and reliability of the rotation curves. 

In Fig.~\ref{pvrc2}, we show the H$\alpha$ and {H{\sc i}}
rotation curves
projected onto the {H{\sc i}} position-velocity diagrams of the galaxies 
of our sample: 
the agreement between the two datasets
is good and, as should be expected, the {H{\sc i}} rotation curves follow the 
shape of the last contours of the
position-velocity diagrams. 
The {H{\sc i}} rotation curves agree with the H$\alpha$ data within 1 $\sigma$ for 
the bulk of the points. In the case of NGC 1090, however, at some radii there is an offset 
of about 10 km s$^{-1}$ between the two data sets: in Fig.~\ref{pvrc2} 
one can see that the H$\alpha$ data lie above the {H{\sc i}} rotation curve; the figure 
shows that it is not 
an error in the {H{\sc i}} data reduction, but rather an intrinsic difference, maybe 
due to non-circular motions associated with the small bar visible in the optical image.
The optical image of ESO 116-G12 looks a bit asymmetric, but the kinematics is not, while
the other galaxies look symmetric in both the kinematics and the photometry, which tells
us that apart from NGC 1090 the effects of possible hidden bars are negligible.
We plot about two points per beam for the {H{\sc i}} rotation curve: this was done in order
to have a comparable number of {H{\sc i}} and H$\alpha$ points. Plotting fewer {H{\sc i}}
points would result in the H$\alpha$ data outnumbering the {H{\sc i}} data and biasing the 
rotation curve fits. 

We also compared the H$\alpha$ data with the
{H{\sc i}} rotation curves derived with two different
methods (Fig.~\ref{rc}): our corrected MET/WAMET method and the first-moment.
The latter yields 
a much worse agreement with the H$\alpha$ rotation curves, 
especially in the inner parts where the profiles
are most asymmetric: in these cases the rotation velocities are severely 
underestimated. This shows the importance of the particular care that
should be taken in deriving rotation curves.

The symmetry and smoothness of the rotation curves
in the selected sample, along with the good agreement 
between the two datasets make these galaxies ideal
laboratories for a dark matter study based on the kinematics: 
this holds in particular because they lack large non-circular motions, a major factor 
of disturbance and uncertainty in this kind of study. 
The agreement between the H$\alpha$ and {H{\sc i}} rotation curves 
also tells us that any systematic errors that might artificially
render the rotation curve shallower (i.e. extinction in the H$\alpha$ data,
bad positioning of the slit, low resolution of the {H{\sc i}} data) are small.

Note that the H$\alpha$ data are absolutely necessary for a mass decomposition such as 
the one presented in the next sections. The uncertainties in the inner parts of the
{H{\sc i}} rotation curves are quite large (typically 10--20 km s$^{-1}$), therefore
nearly all the mass models fitted would be consistent with the data,
disabling us from putting any constraints on the dark matter distribution.
The actual uncertainties on the rotation velocities are probably smaller,
due to the precision than can be reached with the data cube modelling; however,
they are difficult to estimate and rather subjective, therefore we keep
the above estimated errors.

\section{Mass models}
\label{massmodels}

The sample allows us to test in detail different mass models with
various distributions and relative amounts of dark matter.

\subsection {Luminous matter}

In order to study the properties of luminous and dark matter in these galaxies,
we model the circular velocity $V(r)$ in terms of the disc, gas and halo components: 

\begin{equation}
V^2(r)=V^2_{disc}(r)+V^2_{gas}(r)+V^2_{halo}(r).
\end{equation}

$V_{disc}(r)$ is the contribution of the stellar disc, which is scaled
according to the chosen (stellar, I-band) M/L ratio (hereafter 
$\Upsilon_{*}^{\rm I}$). In this way, 
the molecular gas is automatically taken into account,
since it is reasonable to assume that it is distributed like the stellar component
(e.g. \citealt{Cor:00}). 
The available photometry shows that the luminous matter in these galaxies
is distributed like in an exponential thin disc, whose value is given
in Table \ref{physical} \citep*{Ma:92}.
The exception to this is ESO 79-G14, whose profile is significantly 
different from exponential: for this galaxy the actual surface brightness
profile was used.
$\Upsilon_{*}^{\rm I}$ (assumed to be constant
with radius) is a free parameter in each model, however, with the condition 
of being larger than 0.2.

$V_{gas}(r)$ is the contribution of the gaseous disc, derived from the
{H{\sc i}} surface density distribution, scaled up by a factor 1.33
to account for primordial helium; 
the uncertainty of the galaxy distance is irrelevant for
the present study since at no radius the {H{\sc i}} contribution dominates
the kinematics.

$V_{halo}(r)$ is the contribution of the dark matter halo, derived 
under a number of assumptions as discussed below.

\subsection{Dark matter}
\subsubsection{Burkert halo} 

\citet{B:95} proposed an empirical density distribution of
dark matter:

\begin{equation}
\rho_{\rm B}(r)=\frac{\rho_0 r_{core}^3}{(r+r_{core})(r^2+r_{core}^2)}.
\label{burkerteq}
\end{equation}

Here, $\rho_0$ is the central density and $r_{core}$ the core radius.
This halo has a constant density core, 
$\rho_{\rm B}(r) \rightarrow \rho_0$ for $r \rightarrow 0$, and a decrease
of the density as $r^{-3}$ for $r \rightarrow \infty$. 

The halo has two free parameters: 
the core radius $a=r_{core}/r_{opt}$ (i.e., in units of the optical radius, $r_{opt}=3.2r_{d}$) 
and the central density $\rho_0$.

Other cored distributions
were investigated (the pseudo-isothermal halo, \citealt{vA:85}, and
the Universal Rotation Curve halo, \citealt*{PSS:96}, hereafter PSS),
but they yield results which are almost indistinguishable from the
Burkert halo. We include them in Table \ref{tab-param} for completeness. 

\subsubsection{Navarro, Frenk and White (NFW) halo} 

\citet{NFW:96} fitted the outcome of N-body 
CDM simulations with a density distribution given by:

\begin{equation}
\rho_{\rm NFW}(r)=\frac{\rho_s}{(r/r_s)(1+r/r_s)^2}.
\end{equation}

Here, $\rho_s$ and $r_s$ are the characteristic density and the
scale radius of the distribution. 
They are in principle independent,
but recent results (\citealt{Bul:01}, \citealt{We:02})
show a correlation between these parameters. The NFW 
density distribution is then a one-parameter family, namely the
virial mass $M_{vir}$. 
From \citet{We:02} we take the 
relations linking $M_{vir}$ to the concentration parameter $c~(=r_{vir}/r_s)$, 
$r_s$ and $\rho_s$, at redshift $z=0$
and for a Universe with $\Lambda = 0.7$ and $\Omega_0 = 0.3$, starting with 
$M_{vir} \equiv \frac {4}{3} \pi \Delta_{vir} \rho_{c} r_{vir}^3$
(where $\Delta_{vir}$ is the virial overdensity and its value is about 337 at $z=0$,
$\rho_{c}$ is the critical density of the Universe and $r_{vir}$ is the 
virial radius):

\begin{equation}
\label{cmvir}
c \simeq 20 \left( \frac{M_{vir}}{10^{11} {\rm M_{\odot}}} \right)^{-0.13},~~ 
r_s \simeq 5.7 \left( \frac{M_{vir}}{10^{11} {\rm M_{\odot}}} \right)^{0.46}~{\rm kpc}
\end{equation} 

\begin{equation}
\label{rhosnfw}
\rho_s \simeq \frac{101}{3} \frac{c^3}{{\rm ln}(1+c)-\frac{c}{1+c}} \rho_{c}.
\end{equation} 

NFW halo has then a central density cusp, with $\rho_{\rm NFW} 
\propto r^{-1}$ for $r \rightarrow 0$, and a 
profile/amplitude which is controlled by a free parameter $M_{vir}$.

Notice that in principle, 
adiabatic contraction of the primordial dark matter halo due to baryon infall 
should be taken into account, but since the effect is to render the
halo even more concentrated, aggravating thus the known problems of the
NFW haloes, we neglect it.

We constrain the virial halo mass to be $M_{vir}<8 \times 10^{11}$ M$_{\odot}$, in 
that, for a low luminosity spiral,
it must presumably be substantially lower than that of the Milky Way 
and other very luminous galaxies, for which it is
safely estimated : $M_{vir} \simeq 2 \times 10^{12}$ M$_{\odot}$ (\citealt{C:93}, \citealt{W:99}).
This constraint affects only N7339 and ESO 79-G14, due to the relatively limited extension 
of their {H{\sc i}} rotation curves, which prevents to rule out
large $\Lambda$CDM haloes.

\subsubsection{Moore halo}

Recent numerical simulations by \citet{Mo:98} yielded
a more concentrated density profile:

\begin{equation}
\rho_{\rm Moore}(r)=\frac{\rho_s}{(r/r_s)^{1.5}(1+(r/r_s)^{1.5})},
\end{equation}

where $\rho_s$ and $r_s$ are the characteristic density and the
scale radius of the distribution. 
This density 
distribution has an even steeper cusp ($\rho_{\rm Moore} 
\propto r^{-1.5}$ for $r \rightarrow 0$)
than the previous one.
Similarly to the NFW halo, we consider this profile as having only
one free parameter. Following \citet{Mo:99}, we define
$c_{\rm Moore}$ as being 1.8 times smaller
than $c_{\rm NFW}$; it is then derived from Eq. \ref{cmvir}.
For a given virial radius, the scale radius $r_s$ of the
Moore halo will then be 1.8 times larger
than its corresponding quantity for the NFW halo. $\rho_s$ 
can be derived from:

\begin{equation}
\label{rhosmoore}
\rho_s=\frac{101}{2} \frac{c^3}{{\rm ln}(1+c^{1.5})} \rho_{c}.
\end{equation} 

Also in this case we constrain the virial halo mass to be lower than
$8 \times 10^{11}$ M$_{\odot}$.

\subsubsection{{H{\sc i}} scaling}

Early studies of {H{\sc i}} rotation curves \citep{Bo:81}
noted the fact
that the ratio between the {H{\sc i}} surface density and the 
dark matter surface density is approximately constant in the outer 
parts of galaxies (but see \citealt{Cor:00}). This led to the 
hypothesis that dark matter could in some way be associated with 
the {H{\sc i}} disc and distributed in the same manner;
this is what is reasonable to expect in the case of models considering
for instance H$_2$ clumps as a component of dark matter 
\citep*{Pf:94}.
In this case the scaling factor for the {H{\sc i}} contribution
to the rotation curve is a free parameter.

\subsection{MOND}

According to MOND, the law of Modified Newtonian
Dynamics \citep{Mi:83}, there exists a certain acceleration $a_0$ below
which Newton's law of gravity is no longer valid
and the expression 
for the gravitational acceleration reads:

\begin{equation} 
g(r)=[GM(r)a_0]^{1/2}/r,
\end{equation}

where $M(r)$ account for the stellar and gaseous components and
$a_0=1.2 \times 10^{-8}$ cm s$^{-2}$ \citep*{Beg:91}.

\begin{table*}
\centering
\begin{scriptsize}
\centering
\caption[]{Best-fitting results: parameters and associated errors,
reduced $\chi^2$. 
$a=r_{core}/r_{opt}$,
$\rho_0$ is in units of
$10^{-24}$ g cm$^{-3}$ and $M_{vir}$ is in units of $10^{11}$ M$_{\odot}$.}
\vspace{0.3cm}
\label{tab-param}
\begin{tabular} {l c c c c c c} 
\hline
\hline
Halo type \vspace{0.2cm}& Parameter& ESO 116-G12         & ESO 287-G13            & ESO 79-G14             & NGC 1090               & NGC 7339          \\ 
\hline  

\vspace{0.2cm}

Burkert          & $a          $& 0.79$^{+0.21}_{-0.18}$ & 2.29$^{+0.56}_{-0.41}$ & 0.67$^{+0.08}_{-0.07}$ & 0.81$^{+0.23}_{-0.22}$   & 1.60$^{+ \infty}_{-0.79}$ \\ \vspace{0.2cm}
                 & $\rho_0$     & 3.1$^{+2.3}_{-1.2}$    & 0.36$^{+0.08}_{-0.07}$ & 2.2$^{+0.6}_{-0.7}$    & 1.3$^{+1.4}_{-0.5}$     & 2.8$^{+2.7}_{-1.4}$   \\  \vspace{0.2cm}

                 & $\Upsilon_{*}^{\rm I}$ & 0.53$^{+0.24}_{-0.33}$ & 1.83$^{+0.07}_{-0.08}$ & 0.73$^{+0.27}_{-0.25}$ & 1.50$^{+0.24}_{-0.36}$   & 1.84$^{+0.15}_{-0.24}$ \\\  \vspace{0.2cm}
                 &$\chi^2_{red}$& 3.0                    & 1.2                    & 1.1                    & 0.4                      & 1.2                     \\  
\hline
\vspace{0.2cm}
URC              & $a          $& 0.59$^{+0.19}_{-0.12}$ & 2.20$^{+0.42}_{-0.36}$ & 0.67$^{+0.09}_{-0.08}$ & 0.56$^{+0.31}_{-0.16}$    & 1.47$^{+3.94}_{-0.55}$\\ \vspace{0.2cm}
                 & $\rho_0$     & 4.2$^{+2.7}_{-1.9}$    & 0.28$^{+0.06}_{-0.04}$ & 1.5$^{+0.4}_{-0.3}$   & 1.9$^{+2.0}_{-1.1}$         & 2.3$^{+1.5}_{-0.8}$\\ \vspace{0.2cm}
                 & $\Upsilon_{*}^{\rm I}$   & 0.30$^{+0.30}_{-0.10}$ & 1.86$^{+0.06}_{-0.09}$ & 0.90$^{+0.20}_{-0.24}$ & 1.32$^{+0.33}_{-0.28}$ & 1.85$^{+0.14}_{-0.17}$ \\ \vspace{0.2cm}
                 &$\chi^2_{red}$& 2.5                    & 1.2                    & 0.9                    & 0.4                       & 1.1                    \\
\hline
\vspace{0.2cm}
pseudo-isothermal& $a          $& 0.40$^{+0.10}_{-0.12}$ & 1.50$^{+0.29}_{-0.32}$ & 0.42$^{+0.09}_{-0.05}$ & 0.37$^{+0.15}_{-0.13}$     & 1.10$^{+3.61}_{-0.38}$ \\ \vspace{0.2cm}
                 & $\rho_0$     & 3.6$^{+2.3}_{-1.3}$    & 0.30$^{+0.08}_{-0.05}$ & 1.6$^{+0.5}_{-0.5}$    & 1.7$^{+2.1}_{-1.0}$        & 2.3$^{+1.5}_{-0.9}$  \\ \vspace{0.2cm}
                 & $\Upsilon_{*}^{\rm I}$   & 0.50$^{+0.21}_{-0.24}$ & 1.84$^{+0.07}_{-0.09}$ & 0.96$^{+0.24}_{-0.17}$ & 1.50$^{+0.22}_{-0.26}$ & 1.85$^{+0.15}_{-0.17}$ \\ \vspace{0.2cm}
                 &$\chi^2_{red}$& 2.8                    & 1.2                    & 1.3                    & 0.5                        & 1.1  \\
\hline
\vspace{0.2cm}
NFW              &$M_{vir}$  & 1.71$^{+0.10}_{-0.12}$ & 7.59$^{+0.42}_{-0.38}$ & 8.00$^{+0.00}_{-0.35}$& 4.99$^{+0.27}_{-0.26}$     & 8.00$^{+0.00}_{-0.56}$ \\ \vspace{0.2cm}
                 & $\Upsilon_{*}^{\rm I}$   & 0.20$^{+0.01}_{-0.00}$ & 0.71$^{+0.05}_{-0.05}$ & 0.49$^{+0.06}_{-0.04}$ & 0.96$^{+0.08}_{-0.09}$  & 1.32$^{+0.04}_{-0.04}$  \\ \vspace{0.2cm}
                 &$\chi^2_{red}$& 5.2                    & 2.7                    & 7.2                    & 1.1                         & 1.4    \\ 
\hline
\vspace{0.2cm}
Moore            &$M_{vir}$  & 1.44$^{+0.10}_{-0.12}$ & 7.51$^{+0.34}_{-0.41}$ & 6.20$^{+0.81}_{-0.69}$& 4.81$^{+0.27}_{-0.25}$     & 8.00$^{+0.00}_{-1.47}$ \\ \vspace{0.2cm}
                 & $\Upsilon_{*}^{\rm I}$  & 0.20$^{+0.01}_{-0.00}$ & 0.56$^{+0.06}_{-0.06}$ & 0.65$^{+0.11}_{-0.13}$ & 0.91$^{+0.09}_{-0.08}$  & 0.93$^{+0.12}_{-0.04}$  \\ \vspace{0.2cm}
                 &$\chi^2_{red}$& 10.6                    & 6.6                    & 13.0                   & 2.7                       & 5.3  \\
\hline
\vspace{0.2cm}
{H{\sc i}}-scal & scal fact    & 6.49$^{+0.28}_{-0.28}$ & 4.81$^{+0.13}_{-0.14}$ & 7.67$^{+0.45}_{-0.49}$& 6.76$^{+0.23}_{-0.23}$      & 11.63$^{+0.97}_{-1.08}$\\ \vspace{0.2cm}
                 & $\Upsilon_{*}^{\rm I}$   & 1.28$^{+0.06}_{-0.05}$ & 2.14$^{+0.04}_{-0.04}$ & 2.00$^{+0.07}_{-0.06}$ & 2.32$^{+0.07}_{-0.07}$ & 2.12$^{+0.06}_{-0.05}$\\ \vspace{0.2cm}
                 &$\chi^2_{red}$& 18.8                   & 11.1                   & 7.0                   & 2.8                        & 1.1    \\
\hline
\vspace{0.2cm}
MOND             & $\Upsilon_{*}^{\rm I}$   & 0.81$^{+0.02}_{-0.03}$ & 1.48$^{+0.02}_{-0.02}$ & 1.24$^{+0.02}_{-0.02}$ & 1.23$^{+0.02}_{-0.02}$ & 2.15$^{+0.03}_{-0.03}$ \\ \vspace{0.2cm}
                 &$\chi^2_{red}$& 5.2                    & 1.9                    & 2.5                    & 3.4                       & 3.6  \\  
\hline
\end{tabular}
\end{scriptsize}
\end{table*}

\section{Mass modelling results}
\label{bestfits}

The fits were performed by a $\chi^2$-minimisation, considering
both the rotational velocities and their logarithmic gradients 
($\nabla= \frac {d{\rm log} V(r)}{d{\rm log}r}$), 
which bear a crucial information on the matter distribution
in a galaxy (see \citealt{PS:90}). 
The total $\chi^2$ value to be minimised then is
$\chi^2_{tot}=\chi^2_{vel}+\chi^2_{\nabla}$.
It is worthwhile to point out that the $\chi^2$ values should
only be considered as a way to compare the different fits
within the same galaxy, rather than a probability indicator,
because the choice of the error bars is quite subjective and
we plot two points per beam, so the points are not independent; 
the goodness of a particular mass model
is also related to the fraction of observational points that it hits within 1 $\sigma$
as well as the ones that it badly misses. 

\begin{figure*}
  \centerline{\psfig{figure=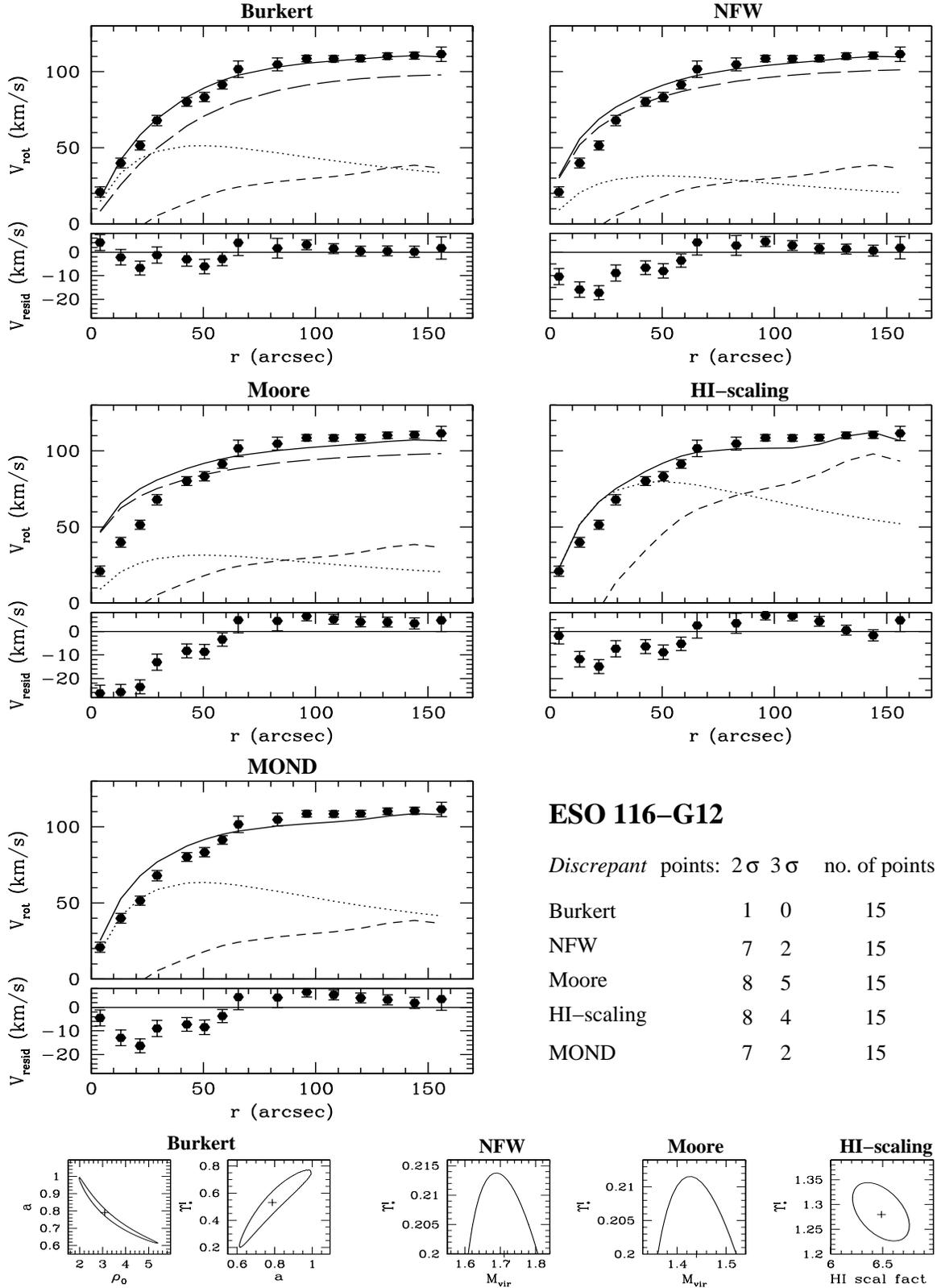,width=157.1mm}}
  \caption{Mass models for the galaxy ESO 116-G12. The solid line represents
the best-fitting model, the long-dashed line is the contribution of the dark matter
halo, and the dotted and short-dashed lines are the contributions
of the stellar and gaseous discs, respectively. 1 kpc corresponds to 
13{\hbox{$.\!\!^{\prime\prime}$}}4.
Below the rotation curves we show the residuals ($V_{obs}~-~V_{model}$).
At the bottom we plot the 1 $\sigma$ probability 
contours in the parameter space, where the crosses indicate the best-fitting
values, $\rho_0$ is in 
units of 10$^{-24}$ g cm$^{-3}$ and $M_{vir}$ is in units of $10^{11}$ M$_{\odot}$.
We also give the number of discrepant points, with residuals
is larger than 2 $\sigma$ and 3 $\sigma$, where $\sigma$ is the observational error
on $V_{obs}(r)$ at radius $r$. 
}
\label{116}
\end{figure*}

\begin{figure*}
  \centerline{\psfig{figure=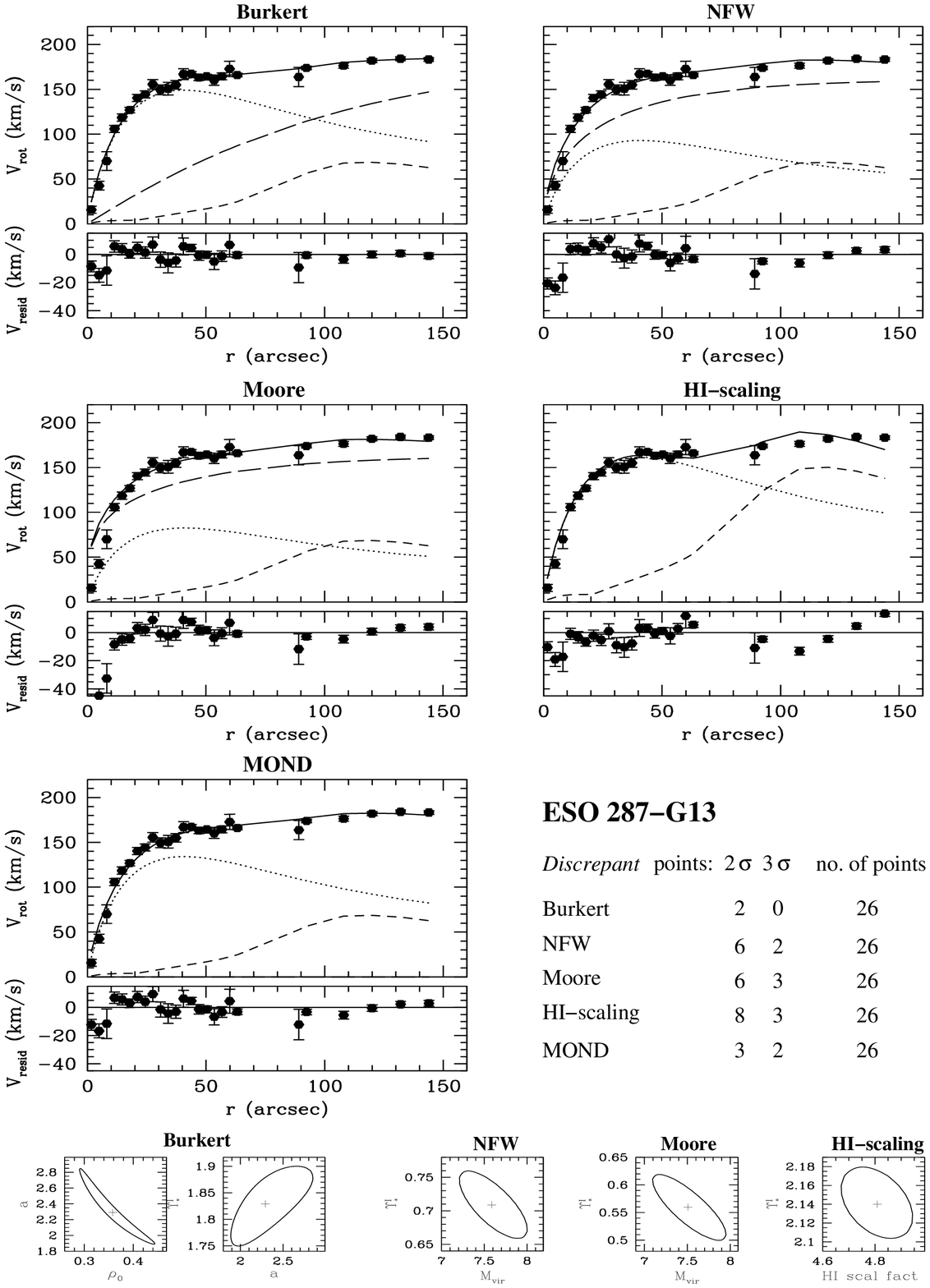,width=157.1mm}}
  \caption{Mass models for the galaxy ESO 287-G13 (see Fig.~\ref{116} for an 
explanation of layout and symbols). 1 kpc corresponds to 
5{\hbox{$.\!\!^{\prime\prime}$}}8.
}
\label{287}
\end{figure*}

\begin{figure*}
  \centerline{\psfig{figure=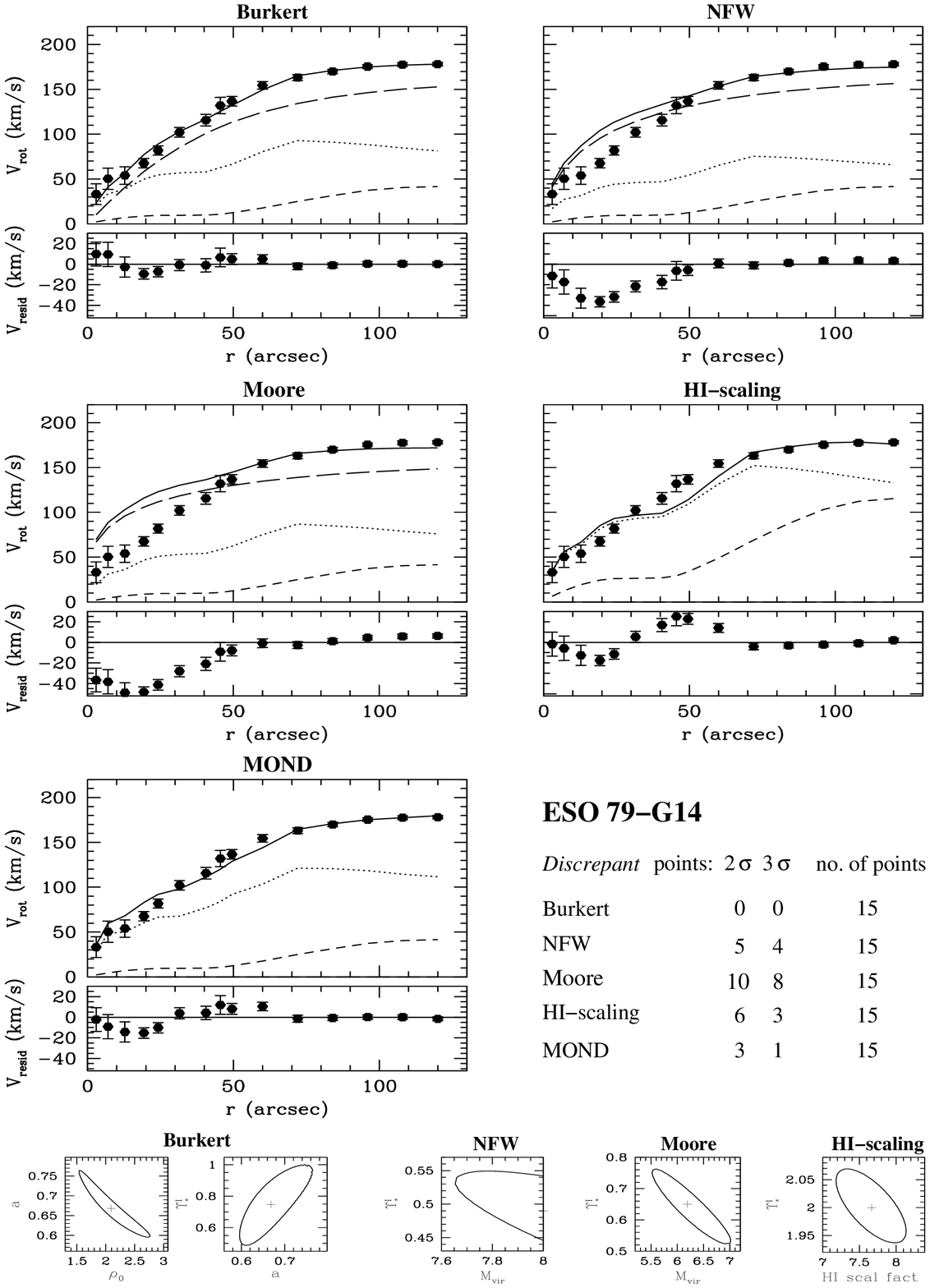,width=157.1mm}}
  \caption{Mass models for the galaxy ESO 79-G14 (see Fig.~\ref{116} for an 
explanation of layout and symbols). 1 kpc corresponds to 
6{\hbox{$.\!\!^{\prime\prime}$}}8.
}
\label{79}
\end{figure*}

\begin{figure*}
  \centerline{\psfig{figure=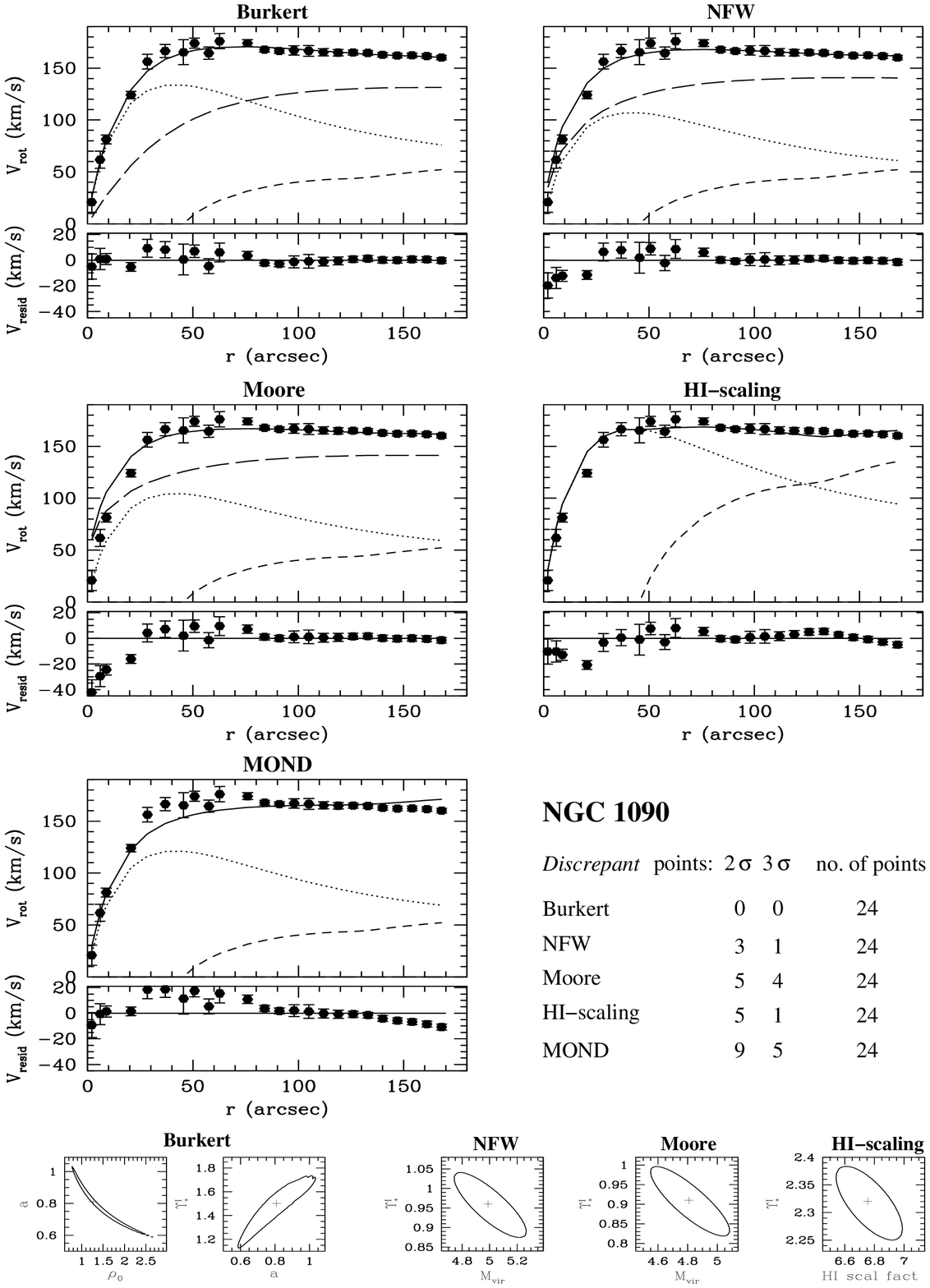,width=157.1mm}}
  \caption{Mass models for the galaxy NGC 1090 (see Fig.~\ref{116} for an 
explanation of layout and symbols). 1 kpc corresponds to 
5{\hbox{$.\!\!^{\prime\prime}$}}7.
}
\label{1090}
\end{figure*}

\begin{figure*}
  \centerline{\psfig{figure=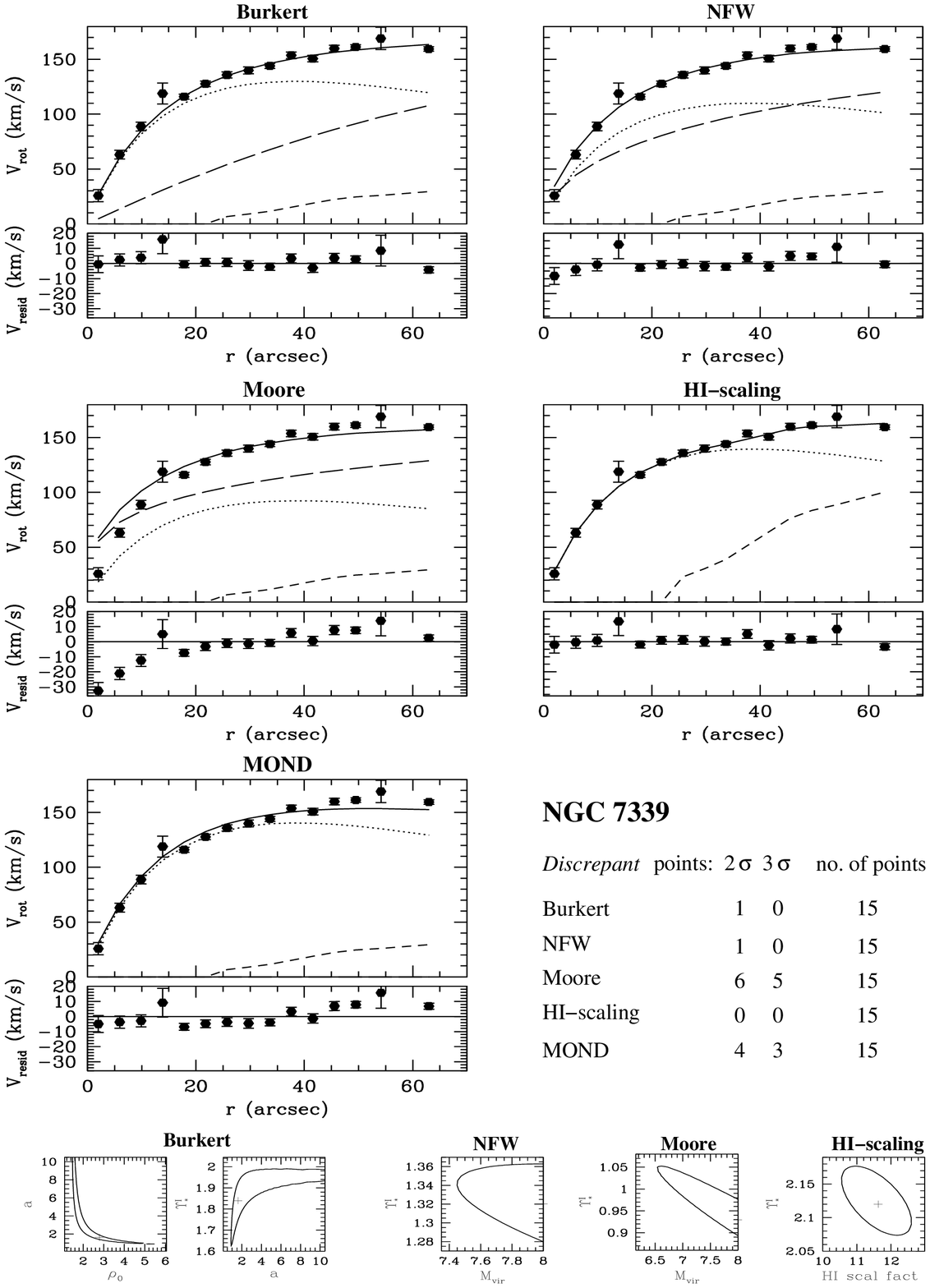,width=157.1mm}}
  \caption{Mass models for the galaxy NGC 7339 (see Fig.~\ref{116} for an 
explanation of layout and symbols). 1 kpc corresponds to 
11{\hbox{$.\!\!^{\prime\prime}$}}6.
}
\label{7339}
\end{figure*}

In Figs. \ref{116} to \ref{7339} we show, for each galaxy, the results of the fits, 
the residuals ($V_{obs}-V_{model}$) of the fits and
the 1 $\sigma$ probability contours in parameter space.
The case of NGC 7339 will be discussed in Appendix A.

\begin{figure}
  \centerline{\psfig{figure=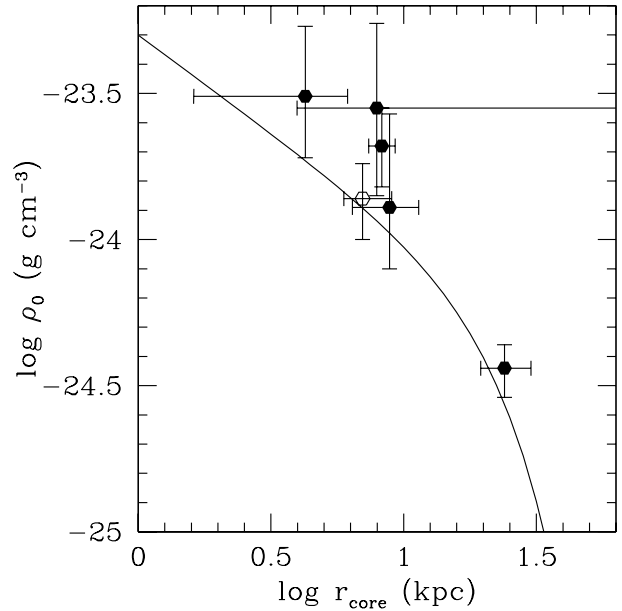,width=85mm}}
  \caption{Relation between the Burkert central density and Burkert core radius;
the solid line is the relation found by \citet{B:95} and extended to spiral
galaxies by \citet{SB:00}: 
$\rho_0= 5\times 10^{-24} r_{core}^{-2/3} e^{-(r_{core}/27)^2}$ g cm$^{-3}$. 
The open circle represents
DDO 47 \citep{Sa:03} and the filled circles are the galaxies of the present sample.
}
\label{bur}
\end{figure}

The Burkert profile --
so as any cored profile -- has the best fits
to the rotation curves, with no systematic deviation from the
observed rotation curves seen in all galaxies. 
None of our $\sim$ 100 data points (considering the five 
galaxies together) is inconsistent with this model, 
having a residual larger than 3 $\sigma$ (where $\sigma$ is the observational error).
The stellar I-band mass-to-light ratios, which lie between
0.5 and 1.8, are consistent with population synthesis models (e.g., \citealt{BdJ:01}).
The core radii are in the range (0.7 -- 2.3)
$\times~ r_{opt}$, and the central densities are between (0.4 --
3) $\times$ 10$^{-24}$ g cm$^{-3}$. 
In Fig. \ref{bur} we plot
the galaxies of our sample in the $\rho_0~-~r_{core}$ plane of \citet{B:95},
slightly adapted to spiral galaxies by \citet{SB:00}:
despite a certain scatter, they roughly follow the relation, which certainly
has an implication for the nature of dark matter. 

The minimum $\chi^2$ values for the NFW haloes
are significantly higher than for the Burkert haloes. The former fail to 
reproduce both the velocities and the shape of the observed rotation curves.
Moreover, there is a systematic effect in the predicted velocities,
in the sense that 
the NFW haloes predict velocities in the central parts
that are too high.
A significant number of data points (9) is totally inconsistent (i.e., residuals
larger than 3 $\sigma$) with 
the fits performed with the NFW halo; if we consider the points with residuals
larger than 2 $\sigma$ their number increases to 22.
An additional problem
of the NFW haloes is 
that the emerging $\Upsilon_{*}^{\rm I}$ 
are generally small and in one case unacceptably low; in this case the extreme  
lower limit of 0.2 is reached. 
We also attempted to reproduce the data by leaving $c$ as a free parameter,
instead of using Eq. \ref{cmvir}: no appreciable improvement of the fits
is found. Moreover,
the values of $c$ and $M_{vir}$
are in some cases completely unrealistic, in particular we obtain very low values
of $c$.

The Moore haloes provide 
even worse results, with very high $\chi^2$ values compared to the cored
haloes, and velocities that are far too high in the inner parts, which
are a consequence of the predicted steep central cusps.
The inability of the Moore haloes to reproduce the observed kinematics
appears also in the large number of points (25) having residuals larger
than 3 $\sigma$. Notice that recent simulations (\citealt{N:03}) predict, 
at the innermost resolved radius, a slope of 1.2 (approximately intermediate
between the NFW and the Moore profile) in the radial density 
profile of a typical galaxy and that they rule out the Moore profile.

The {H{\sc i}}-scaling and the MOND models
give reasonable results in a few cases, but 
with some decisive counter-examples:
we have a significant number of points that are inconsistent 
with the fits in both cases. In particular, the present data have a resolution
such that MOND is not able to fit them.
The values of the {H{\sc i}} scaling factors are consistent with
previous studies (\citealt{Ho:01}).
Leaving the MOND acceleration parameter $a_0$ as a free parameter 
only slightly improves the quality of the fits: they are always
significantly worse than in case of the cored haloes. The same holds
if we leave a reasonable (25\%) uncertainty on the adopted distance.

\section{Conclusions}

We have presented an analysis of {H{\sc i}} data cubes of a sample of
five low-luminosity spiral galaxies that are used to derive
the distribution of dark matter within these galaxies, along with
available H$\alpha$ and I-band photometric data. 

We found a new method (the MET/WAMET method) for deriving the {H{\sc i}} rotation curves, 
which was needed to take into account a) a position angle that varies
with radius (this is needed as the tilted-ring modelling of the velocity field could be used
in only one case), and b) the asymmetry
of the velocity profiles, because the galaxies of our sample have a high inclination,
and projection effects yield asymmetric velocity profiles. Corrections
were applied in order to account for any effects that artificially broaden 
the {H{\sc i}} profiles.

The rotation curves derived in this way passed a crucial test,
i.e. the construction of model data cubes using these rotation curves as an
input. In some cases the {H{\sc i}} rotation curves derived from MET/WAMET required a small
modification, mostly in the inner parts. After these small corrections the
model data cubes are in very good agreement with the observations,
contrary to model data cubes constructed using rotation curves derived
from other ``classical'' methods. 
Our final {H{\sc i}} rotation curves are consistent with the H$\alpha$ data,
but up to three times more extended.

In galaxies well
suited for inferring the distribution of dark matter, 
we were able to trace the total potential via the rotation velocity and 
to derive the density distribution of dark matter. 

From the analysis of combined H$\alpha$ and {H{\sc i}}
rotation curves 
we reach the following conclusions:

\begin{itemize} 
\item[$\bullet$]{Cored haloes uniquely fit the rotation curves, with 
core radii of order $r_{opt}$ and central densities of order 
$10^{-24}$ g cm$^{-3}$.}

\item[$\bullet$]{NFW haloes cannot represent the dark matter haloes
around the galaxies of our sample: they fit badly and
a significant number of data points is inconsistent with the fits.}

\item[$\bullet$]{The situation is even worse for the Moore model,
as implied by its steeper inner profile.} 

\item[$\bullet$] {
There are cases in which for radii in the range $(1-4)r_d$ 
the Burkert and NFW halo rotation curves are similar; 
this may explain why in the literature cases can be found 
(e.g. \citealt{J:03}) in which
both models fit equally well a rotation curve of small extension
and limited resolution.}

\end{itemize}

We thus reach the conclusion that the galaxies
of the present sample are uniquely successfully fitted by cored
haloes, with a core size comparable to the optical radius.
This suggests the existence of a well-defined scale length in 
dark matter haloes, linked to the luminous matter, which is totally 
unexpected in the framework of CDM theory (see \citealt{SB:00}).

\section*{Acknowledgements}
The authors wish to thank the referee, Albert Bosma, for his
comments that improved the outline and the quality of the paper.
We thank Thomas Fritz and Christian Br\"uns 
for help with the observations.
GG and DV are grateful for financial support from the {\it 
Deutsche Forschungsgemeinschaft} under number GRK 118 ``The Magellanic System, 
Galaxy Interaction and the Evolution of Dwarf Galaxies''.
UK is very grateful to the kind hospitality as SISSA
during several visits.

\appendix

\section {The case of NGC 7339}

The galaxy NGC 7339, due to the small extent of the kinematical data and
the value of its dark mass, shows a degeneracy between the predicted Burkert 
and NFW halo rotation curves. As can be seen in Fig. \ref{7339} and in Table \ref{tab-param},
both models provide equally good fits to the rotation curve. 
This is not due to the fitting procedure but it derives from the fact that
both models have similar predictions.  We treat in detail this case to show that $\Lambda$CDM haloes, for
certain virial masses and in certain ranges of radii has a radial profile that meets that 
of the Universal Rotation Curve (see PSS).

\begin{figure*}
  \centerline{\psfig{figure=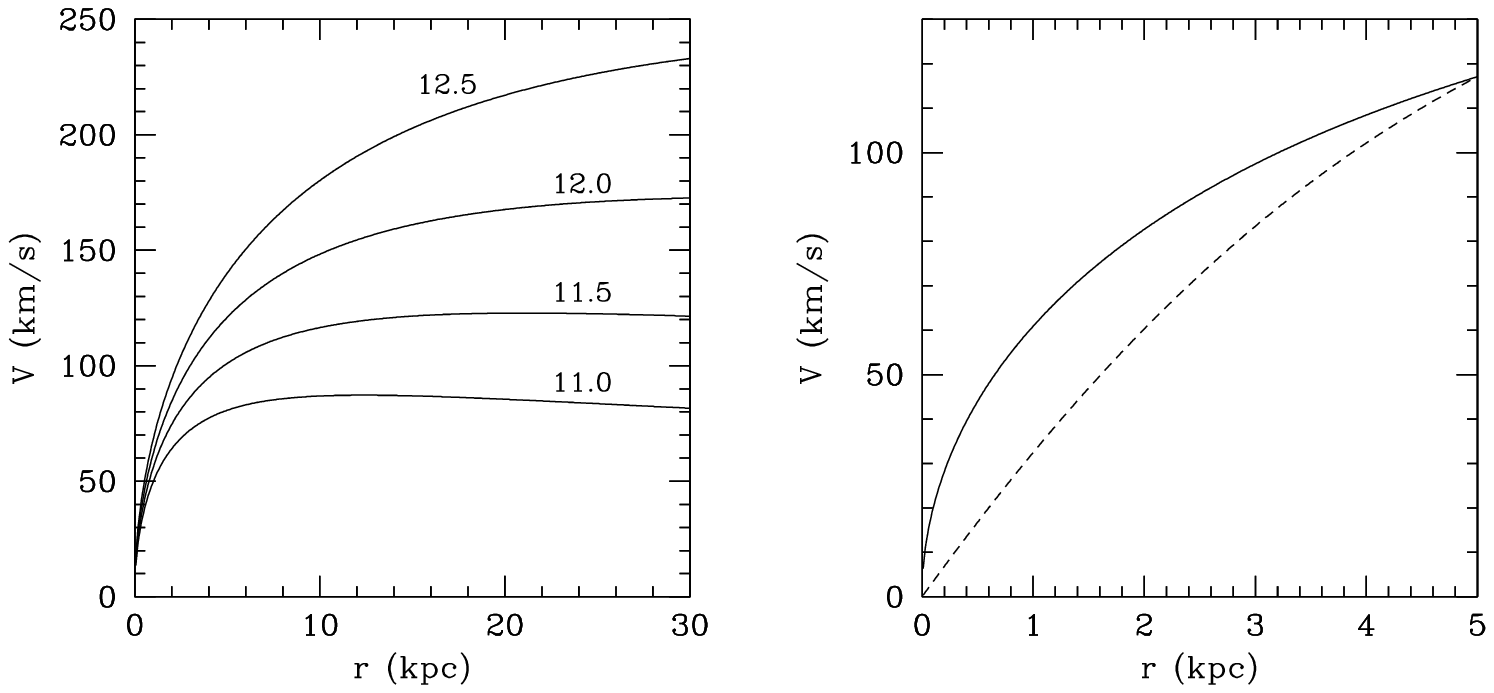,width=170mm}}
  \caption{
Left: NFW halo rotation curves for galactic haloes; the values
of log($M_{vir}/$M$_{\odot}$) are indicated on the plot. Notice the
very large differences with respect to the Universal Rotation Curve (Fig. 6 of PSS).
Right: NFW halo rotation curve (solid line) for an object with $M_{vir}=8 \times 10^{11} 
$M$_{\odot}$ and a Burkert halo rotation curve with $r_{core}=6$ kpc and $\rho_0=4.5 \times 10^{-24}$
g cm$^{-3}$, for radii up to 5 kpc. 
}
\label{halos}
\end{figure*}

First, let us notice that from Table \ref{physical} that the rotation curve of
this galaxy has a small spatial extent, i.e. about 5 kpc. 
Then in Fig. \ref{halos} (left panel) we show the NFW halo rotation curve,
very different in general from the derived actual haloes (\citealt{SB:00}), for objects having
a number of virial masses, ranging from $1 \times 10^{11}$ M$_{\odot}$
to $3 \times 10^{12}$ M$_{\odot}$. 
However, as shown in the right panel of Fig. \ref{halos}, for virial masses close to
about $1 \times 10^{12}$ M$_{\odot}$, and for radii up to 5 kpc, the
NFW and Burkert halo rotation curves almost coincide;
this similarity, combined with the uncertainty on $\Upsilon_{*}^{\rm I}$, is the reason for the equal
quality of the two fits.

With the data used here we cannot probe the regions of NGC 7339 where the difference is noticeable.
Therefore, for galaxies having $V(r_{opt})$ approximately in the range $120 - 170$ km s$^{-1}$
and for the radial range $1 < r/r_d < 4$,
in order to distinguish between the two cases, we need kinematical
information at larger radii.
At smaller and larger $V(r_{opt})$ the two profiles are intrinsically much more different and
the comparison is easier (see PSS).

\end{document}